\documentclass[10pt]{amsart}
\usepackage{amsmath,amsfonts,amscd,amssymb,epsf} 
\newtheorem{theorem}{Theorem}[section]
\newtheorem{proposition}[theorem]{Proposition}
\newtheorem{lemma}[theorem]{Lemma}
\newtheorem{corollary}[theorem]{Corollary}
\theoremstyle{definition}
\newtheorem{definition}[theorem]{Definition}

\theoremstyle{remark} \newtheorem{remark}[theorem]{Remark}
\numberwithin{equation}{section}

 \DeclareMathOperator{\id}{id}
\DeclareMathOperator{\Id}{Id}

\DeclareMathOperator{\Tr}{Tr} \DeclareMathOperator{\Diff}{Diff}
\DeclareMathOperator{\Mob}{M\ddot{o}b} 
 
\DeclareMathOperator{\sgn}{sgn} 
 \DeclareMathOperator{\PSU}{PSU}

\DeclareMathOperator{\Mobius}{M\ddot{o}bius}
\newcommand{\Del}{\mathbb{D}}

\newcommand{\Z}{{\mathbb{Z}}}

\newcommand{\R}{{\mathbb{R}}}
\newcommand{\C}{{\mathbb{C}}}
\newcommand{\bk}{\backslash}
\newcommand{\pa}{\partial}

\newcommand{\ov}{\overline}

\newcommand{\z}{\bar{z}}

\begin{document}

\title[ Universal Index Theorem on $\Mob(S^1)\bk S^1\bk\Diff_+(S^1)$]
{  Universal Index Theorem on $\Mob(S^1)\bk\Diff_+(S^1)$
}\author{Lee-Peng Teo}\address{Faculty of Information Technology,
Multimedia University, Jalan Multimedia, Cyberjaya, 63100, Selangor
Darul Ehsan, Malaysia}\email{lpteo@mmu.edu.my}
\date{\today}

\begin{abstract}By conformal welding, there is a pair of univalent functions $(f,g)$
associated to every point of the complex K\"ahler manifold
$\Mob(S^1)\bk\Diff_+(S^1)$. For every integer $n\geq 1$, we
generalize the definition of Faber polynomials to define some
canonical bases of holomorphic $1-n$ and $n$ differentials
associated to the pair $(f,g)$. Using these bases, we generalize the
definition of Grunsky matrices to define matrices whose columns are
the coefficients of the differentials with respect to standard bases
of differentials on the unit disc and the exterior unit disc. We
derive some identities among these matrices which are reminiscent of
the Grunsky equality. By using these identities, we showed that we
can define the Fredholm determinants of the period matrices of
holomorphic $n$ differentials $N_n$, which are the Gram matrices of
the canonical bases of holomorphic $n$-differentials with respect to
the inner product given by the hyperbolic metric. Finally we proved
that $\det N_n =(\det N_1)^{6n^2-6n+1}$ and $\pa\bar{\pa}\log\det
N_n$ is $-(6n^2-6n+1)/(6\pi i)$ of the Weil-Petersson symplectic
form.
\end{abstract}
\maketitle

\section{Introduction}

  This paper can be considered as a sequel to our paper \cite{TT1},
in which we study the properties of the Weil-Petersson metric on the
universal Teichm\"uller space. However, a fundamental difference is
that in this paper we no longer work with the entire universal
Teichm\"uller space, but instead we work with its subspace
$\Mob(S^1)\bk\Diff_+(S^1)$, which corresponds to smooth
($C^{\infty}$) mappings. As a result, we make no use of any
quasi-conformal mapping theories, and we hope that this paper may be
more accessible to people working in theoretical physics.

Let $\Del$ and $\Del^*$ denotes the unit disc and its exterior
respectively. In \cite{K}, it was shown that  every element
$\gamma\in S^1\bk\Diff_+(S^1)$ is associated with a unique pair of
univalent functions $f:\Del\rightarrow \C$ and $g:\Del^*\rightarrow
\hat{\C}$ such that $\gamma=g^{-1}\circ f$, $f(0)=0,
g(\infty)=\infty$ and $f'(0)=1$. The (modified)\footnote{In order to
conform with our later definitions, we modify slightly the
definition, which differs from the usual one by some
constants.}Faber polynomials $u[0]_k(w)$ and $v[0]_k(w)$, $k\geq 1$
of $(f,g)$ (see e.g. \cite{Pom, Duren, Teo2}) is defined by
\begin{align}\label{eq65}
u[0]_k(w) = &\frac{1}{\sqrt{\pi k}}(g^{-1}(w))_{\geq 1},
\hspace{1cm}v[0]_k(w)=\frac{1}{\sqrt{\pi k}}(f^{-1}(w))_{\leq -1}.
\end{align}They can be encoded in \begin{align*}
\frac{g'(z)w}{(g(z)-w)g(z)}=& \sum_{k=1}^{\infty}u[0]_k(w)\sqrt{\pi k} z^{-k-1},\\
\frac{f'(z)}{f(z)-w } =& -\sum_{k=1}^{\infty}v[0]_k(w) \sqrt{\pi
k}z^{k-1}.
\end{align*}Define the Grunsky coefficients $b_{mn}$ of $(f,g)$
\cite{Pom, Duren, Teo2}  by
\begin{align*}
\log \frac{g(z)-g(\zeta)}{z-\zeta} =
&b_{0,0}-\sum_{m=1}^{\infty}\sum_{n=1}^{\infty}
b_{mn}z^{-m}w^{-n},\\
\log\frac{g(z)-f(\zeta)}{z}=&b_{0,0}
-\sum_{m=1}^{\infty}\sum_{n=0}^{\infty} b_{m,-n}z^{-m}\zeta^n,\\
\log
\frac{f(z)-f(\zeta)}{z-\zeta}=&b_{0,0}-\sum_{m=0}^{\infty}\sum_{n=0}^{\infty}
b_{-m,-n} z^{m}\zeta^n,
\end{align*}and for $m\geq 0$, $n\geq 1$, $b_{-m, n}=b_{m,-n}$. Then
\begin{align*}
u[0]_k (f(z)) =& \sum_{l=1}^{\infty}
\sqrt{kl}b_{-l,k}\frac{1}{\sqrt{\pi l}}z^{l},\\
u[0]_k(g(z))=& \frac{1}{\sqrt{\pi k}}
z^{k}-\sqrt{\frac{k}{\pi}}b_{0,k}+\sum_{l=1}^{\infty}\sqrt{kl}b_{l,k}\frac{1}{\sqrt{\pi
l}}z^{-l},\\
v[0]_k(f(z))=& \frac{1}{\sqrt{\pi k}}
z^{-k}+\sqrt{\frac{k}{\pi}}b_{0,-k}+\sum_{l=1}^{\infty}\sqrt{kl}b_{-l,-k}\frac{1}{\sqrt{\pi
l}}z^{l},\\
v[0]_k (g(z)) =& \sum_{l=1}^{\infty}
\sqrt{kl}b_{l,-k}\frac{1}{\sqrt{\pi l}}z^{-l}.
\end{align*} The Grunsky matrices $A[0], B[0],C[0], D[0]$ are  semi-infinite matrices defined by
\begin{align*}
A[0] =& ( \sqrt{kl} b_{-l,k})_{l,k\geq 1},\hspace{1.2cm}B[0] = (
\sqrt{kl} b_{l,k})_{l,k\geq 1},\\
C[0] =& ( \sqrt{kl} b_{-l,-k})_{l,k\geq 1},\hspace{1cm}D[0] = (
\sqrt{kl} b_{l,-k})_{l,k\geq 1}.
\end{align*}Grunsky equality \cite{Hummel} (or see \cite{TT1}) says that for any complex numbers $\lambda_k$, $k=
\pm 1, \ldots, \pm m$,
\begin{align*}
\sideset{}{^{\prime}}\sum_{k=-\infty}^{\infty}\left|\sideset{}{^{\prime}}\sum_{l=-m}^m\sqrt{|kl|}b_{kl}\lambda_l\right|^2
=\sideset{}{^{\prime}}\sum_{k=-m}^m |\lambda_k|^2.
\end{align*}In other words,
\begin{align}\label{eq83}
\begin{pmatrix}B[0]& D[0]\\
A[0]& C[0] \end{pmatrix}\begin{pmatrix}B[0]^*& A[0]^*\\
D[0]^*& C[0]^* \end{pmatrix}=\Id.
\end{align}In particular,
\begin{align*}
A[0]A[0]^* +C[0]C[0]^*=\Id, \hspace{1cm}D[0]D[0]^*+B[0]B[0]^*=\Id,
\end{align*}which implies that both $A[0]$ and $D[0]$ define bounded
 operators on $\ell^2$ with norm less than or equal to one. In \cite{TT1}, we defined a basis for $A_{1,2}(\Omega)$ and
$A_{1,2}(\Omega^*)$ -- the Hilbert spaces of square-integrable
holomorphic one-forms on $\Omega=f(\Del)$ and $\Omega^*=g(\Del^*)$
-- by
\begin{align}&\label{eq66} \left\{U[1]_k (w)\,:\, U[1]_k=u[0]_k'(w)=\sqrt{\frac{k}{\pi}}\Bigl(g^{-1}(w)^{k-1}(g^{-1})'(w)\Bigr)_{
\geq 0}, \,k\geq 1\right\},\\&\left\{ V[1]_k(w)\,:\,
V[1]_k(w)=-v[0]_k'(w)=\sqrt{\frac{k}{\pi}}\Bigl(f^{-1}(w)^{-k-1}(f^{-1})'(w)\Bigr)_{
\leq -2},\, k\geq 1\right\}.\nonumber
\end{align}  Then
\begin{align*}
U[1]_k(f(z))f'(z)=& \sum_{l=1}^{\infty}
\sqrt{kl} b_{-l,k} \sqrt{\frac{l}{\pi}}z^{l-1},\\
U[1]_k(g(z))g'(z)=& \sqrt{\frac{k}{\pi}}z^{k-1} -
\sum_{l=1}^{\infty}
\sqrt{kl} b_{l,k} \sqrt{\frac{l}{\pi}}z^{-l-1},\\
V[1]_k(f(z))f'(z)=&\sqrt{\frac{k}{\pi}}z^{-k-1}- \sum_{l=1}^{\infty}
\sqrt{kl} b_{-l,-k}
\sqrt{\frac{l}{\pi}}z^{l-1},\\
V[1]_k(g(z))g'(z)=& \sum_{l=1}^{\infty} \sqrt{kl} b_{l,-k}
\sqrt{\frac{l}{\pi}}z^{l-1}.
\end{align*}The  matrices $A[1], B[1],C[1], D[1]$ are defined by
\begin{align*}
A[1] =& ( \sqrt{kl} b_{-l,k})_{l,k\geq 1},\hspace{1.5cm}B[1] = (
-\sqrt{kl} b_{l,k})_{l,k\geq 1},\\
C[1] =& ( -\sqrt{kl} b_{-l,-k})_{l,k\geq 1},\hspace{1cm}D[1] = (
\sqrt{kl} b_{l,-k})_{l,k\geq 1},
\end{align*} so that their columns correspond to the coefficients of
$U[1]_k(f(z))f'(z)$, $U[1]_k(g(z))g'(z)$, $V[1]_k(f(z))f'(z)$ and
$V[1]_k(g(z))g'(z)$ with respect to the standard bases $$\left\{
\sqrt{\frac{k}{\pi}}z^{k-1}\,:\;k\geq 1\right\}
\hspace{0.5cm}\text{and}\hspace{0.5cm}\left\{
\sqrt{\frac{k}{\pi}}z^{-k-1}\,:\;k\geq 1\right\}$$ of
$A_{1,2}(\Del)$ and $A_{1,2}(\Del^*)$. Obviously, $A[1]=A[0]$,
$B[1]=-B[0]$, $C[1]=-C[0]$ and $D[1]=D[0]$. The Gram matrices of the
bases $\{U[1]_k\,:\, k\geq 1\}$ and $\{ V[1]_k\,:\,k\geq 1\}$ with
respect to the inner product is given by
$$(N_1(\Omega)_{lk})_{l,k\geq 1}=\left(\iint\limits_{\Omega}U[1]_l(w) \ov{U[1]_k(w)}d^2w\right)
_{l,k\geq 1}=A[1]^T\ov{A[1]}=D[1]D[1]^*$$and
$$(N_1(\Omega^*)_{l,k})_{l,k\geq 1}=\left(\iint\limits_{\Omega^*}V[1]_l(w) \ov{V[1]_k(w)}d^2w\right)
_{l,k\geq 1}=D[1]^T\ov{D[1]}=A[1]A[1]^*$$ respectively. We call
$N_1(\Omega)$ and $N_1(\Omega^*)$ the period matrices of holomorphic
one-forms.

Define the kernels
\begin{align*}\mathcal{A}[1](z,w)=&\frac{1}{\pi}\frac{f'(z)g'(w)}{(f(z)-g(w))^2}=
\frac{1}{\pi} \sum_{k=1}^{\infty}\sum_{l=1}^{\infty} klb_{-l,k}
z^{l-1}w^{-k-1} , \\
\mathcal{B}[1](z,w)
=&\frac{1}{\pi}\left(\frac{g'(z)g'(w)}{(g(z)-g(w))^2}-\frac{1}{(z-w)^2}\right)
= -\frac{1}{\pi} \sum_{k=1}^{\infty}\sum_{l=1}^{\infty} klb_{l,k}
z^{-k-1}w^{-l-1}, \\\mathcal{C}[1](z,w)
=&\frac{1}{\pi}\left(\frac{f'(z)f'(w)}{(f(z)-f(w))^2}-\frac{1}{(z-w)^2}\right)
= -\frac{1}{\pi} \sum_{k=1}^{\infty}\sum_{l=1}^{\infty} klb_{-l,-k}
z^{k-1}w^{l-1},\\
\mathcal{D}[1](z,w)=&\frac{1}{\pi}\frac{g'(z)f'(w)}{(g(z)-f(w))^2}=
\frac{1}{\pi} \sum_{k=1}^{\infty}\sum_{l=1}^{\infty} klb_{l,-k}
z^{-l-1}w^{k-1}.
\end{align*}They define operators $\mathcal{A}[1]:
\ov{A_{1,2}(\Del^*)}\rightarrow A_{1,2}(\Del)$, $\mathcal{B}[1]:
\ov{A_{1,2}(\Del^*)}\rightarrow A_{1,2}(\Del^*)$, $\mathcal{C}[1]:
\ov{A_{1,2}(\Del)}\rightarrow A_{1,2}(\Del)$ and $\mathcal{D}[1]:
\ov{A_{1,2}(\Del)}\rightarrow A_{1,2}(\Del^*)$ by
\begin{align*}
\left(\mathcal{A}[1]\phi\right)(z)=\iint\limits_{\Del}
\mathcal{A}[1](z,w) \ov{\phi(w)}d^2w
\end{align*}and similarly for $\mathcal{B}[1], \mathcal{C}[1]$ and
$\mathcal{D}[1]$. With respect to the standard bases of
$A_{1,2}(\Del)$ and $A_{1,2}(\Del^*)$ , their matrices are given by
$A[1]$, $B[1]$, $C[1]$ and $D[1]$ respectively. We showed in
\cite{TT1} that for $\gamma\in \Mob(S^1)\bk\Diff_+(S^1)$,
$\mathcal{B}[1]$ and $\mathcal{C}[1]$ are Hilbert-Schmidt operators.
Therefore by the Grunsky equality \eqref{eq83}, the Fredholm
determinants of the period matrices $A[1]A[1]^*$ and $D[1]D[1]^*$
are well defined. We defined
$\mathfrak{F}_1:\Mob(S^1)\bk\Diff_+(S^1)\rightarrow\R$ by
\begin{align*}
\mathfrak{F}_1 (\gamma)=\log\det (A[1]A[1]^*)=\log \det
(D[1]D[1]^*).
\end{align*}Since $A[1]A[1]^*$ is the matrix of the operator
$\mathcal{A}[1]\mathcal{A}[1]^*$, $\mathfrak{F}_1$ can also be
interpreted as
\begin{align*}
\mathfrak{F}_1=\log \prod_{l=1}^{\infty} \lambda_l,
\end{align*}where $1\geq \lambda_1 \geq \lambda_2\geq \ldots$ are the
eigenvalues of the operator $\mathcal{A}[1]\mathcal{A}[1]^*$. In
this aspect, this function has been considered in \cite{Schiffer1}.

In \cite{Nag92} (see also \cite{NS}), Nag considered a period
mapping on $\Mob(S^1)\bk\Diff_+(S^1)$ in the following way. For any
$\gamma\in \Mob(S^1)\bk\Diff_+(S^1)$ and $k\neq 0$, let
\begin{align*}
\frac{1}{\sqrt{\pi |k|}}\left(\gamma(z)^k -\frac{1}{2\pi
i}\oint_{S^1}\gamma(\zeta)^k \frac{d\zeta}{\zeta}\right)=\sum_{l\in
\Z} \Pi[0]_{lk}\frac{1}{\sqrt{\pi |l|}}z^{l}
\hspace{1cm}\text{for}\;\;z=e^{i\theta}\in S^1,
\end{align*}and define
\begin{align*}
\Pi_1[0]=(\Pi[0]_{lk})_{l,k\geq 1},
\hspace{2cm}\Pi_2[0]=(\Pi[0]_{-l,k})_{l,k\geq 1}.
\end{align*}They satisfy the following identity:
\begin{align}\label{eq60}
\begin{pmatrix} \Pi_1[0]& \ov{\Pi_2[0]}\\\Pi_2[0]&\ov{\Pi_1[0]}\end{pmatrix}
\begin{pmatrix} \Pi_1[0]^* &
-\Pi_2[0]^*\\-\ov{\Pi_2[0]}^*&\ov{\Pi_1[0]}^*\end{pmatrix}=\Id.
\end{align}Nag defined the period mapping  by
$$\gamma\mapsto\ov{\Pi_2[0]}\,\ov{\Pi_1[0]}\,^{-1}.$$We proved that
(\cite{TT1}) \begin{align*} A[0]=& (\Pi_1[0]^*)^{-1}, \hspace{2cm}
B[0]=-\Pi_2[0]^T(\Pi_1[0]^* )^{-1},\\
C[0]=&\ov{\Pi_2[0]}\,\ov{\Pi_1[0]}\,^{-1},\hspace{1.6cm}
D[0]=\ov{\Pi_1[0]}\,^{-1},
\end{align*}and therefore the period mapping can be equivalently
defined by
\begin{align*}
\gamma\mapsto C[\gamma;0],
\end{align*}which is the definition given by Kirillov and Yuriev
\cite{KY2}, and \eqref{eq60} is equivalent to the Grunsky equality
\eqref{eq83}.

Given a smooth curve $\gamma_t\in \Mob(S^1)\bk\Diff_+(S^1)$ where
$\gamma_0=\gamma$, it defines a tangent vector at  $\gamma$ in the
following way. Let $\mathrm{u}_t =\gamma_t\circ \gamma^{-1}$ and
$$\left.\frac{d\mathrm{u}_t}{dt}\right|_{t=0}(z)=\sum_{k\in\Z}c_k z^{k+1}, \hspace{1cm}c_{-k}=-\bar{c}_k,\,z\in S^1.$$
It corresponds to the holomorphic and anti-holomorphic tangent
vectors
\begin{align*}
\mathrm{v}(z) = \sum_{k=2}^{\infty} c_k z^{k+1}
\hspace{1cm}\text{and}\hspace{1cm}\bar{\mathrm{v}}(z)=\ov{\mathrm{v}(z)}=\sum_{k=2}^{\infty}\bar{c}_k
\z^{k+1}.
\end{align*}
 In \cite{TT1}, we showed that the partial derivative of the function
 $\mathfrak{F}_1$ is given by
\begin{align*}
\pa\mathfrak{F}_1(\mathrm{v}) =\frac{1}{12\pi i} \oint_{S^1}
\mathcal{S}(g)(z)\mathrm{v}(z)dz,
\end{align*}where $\mathcal{S}(g)(z)$ is the
Schwarzian derivative of $g$. On the other hand, we proved that the
function $S:\Mob(S^1)\bk\Diff_+(S^1)\rightarrow \R$ defined by
$$S(\gamma) = \iint\limits_{\Del}\left|\frac{f^{\prime\prime}(z)}{f'(z)}\right|^2d^2z
+\iint\limits_{\Del^*}\left|\frac{g^{\prime\prime}(z)}{g'(z)}\right|^2d^2z-4\pi\log|g'(\infty)|$$
satisfies $$\pa S(\mathrm{v}) =i \oint_{S^1}
\mathcal{S}(g)(z)\mathrm{v}(z)dz $$ and
\begin{align*}
\pa\bar{\pa}S(\mathrm{v}, \bar{\mathrm{v}}) =\Vert
\mathrm{v}\Vert_{WP}^2=2\pi\sum_{k=2}^{\infty}(k^3-k) |c_k|^2.
\end{align*}The  implies that $S$ is a
Weil-Petersson potential of $\Mob(S^1)\bk\Diff_+(S^1)$ and
\begin{align}\label{eq110}
\det N_1(\Omega)=\det A[1]A[1]^* =
\exp\left(-\frac{1}{12\pi}S\right).
\end{align}

In \cite{MT2}, inspired by the work of \cite{MT1}, we showed that
for a pair of Riemann surfaces $X$ and $Y$ of genus $g$ which are
 uniformized simultaneously by a quasi-Fuchsian group $\Gamma$,
\begin{align}\label{eq61}
\frac{\det\Delta_n}{\det N_n} =\exp\left(
\frac{6n^2-6n+1}{12\pi}S_{QF}\right)|F(n)|^2, \hspace{1cm}n\geq 2,
\end{align}where $\Delta_n$ is the $n$-Laplacian of the pair
$(X,Y)$ , $N_n$ is the Gram matrix of a basis of holomorphic
$n$-differentials with respect to the inner product induced by
hyperbolic metrics, $S_{QF}$ is a Weil-Petersson potential of the
quasi-Fuchsian deformation space and $F(n)$ is a function defined by
the group elements of $\Gamma$. This formula is the anti-derivative
of the local index theorem (see e.g. \cite{Tz91}) which states that
\begin{align}\label{eq81}
\pa\bar{\pa}\log\det\Delta_n-\pa\bar{\pa}\log\det N_n
=\frac{6n^2-6n+1}{6\pi i}\omega_{WP},
\end{align}where $\omega_{WP}$ is the symplectic two-form
corresponding to the Weil-Petersson metric. In physics notation, the
term $$\frac{6n^2-6n+1}{6\pi i}\omega_{WP}$$ here is the anomaly
term. Since pairs of Riemann surfaces $(X,Y)$ simultaneously
uniformized by quasi-Fuchsian groups and the homogeneous space
$\Mob(S^1)\bk\Diff_+(S^1)$ both sit inside the universal
Teichm\"uller space, it is natural to look for the generalization of
the index theorem \eqref{eq81} and the factorization formula
\eqref{eq61} to $\Mob(S^1)\bk\Diff_+(S^1)$, which is the main
question addressed in this paper. In fact, \eqref{eq110} is the
generalization we are looking for when $n=1$.

To prove the formula \eqref{eq61}, one of the necessary ingredients
is to consider a basis of holomorphic $n$-differentials of $X$ and
$Y$ defined by the Bers integral operator \cite{Bers} $K[n] :
\ov{A_{n,2}(Y)}\rightarrow A_{n,2}(X)$ :
\begin{align*} (K[n]\phi)(z) =&\iint\limits_{Y}
K[n](z,w)\ov{\phi(w)}\rho(w)^{1-n}d^2w,
\end{align*} where with the coordinates
provided by the covering maps $J_1: \Omega\rightarrow X$,
$J_2:\Omega^*\rightarrow Y$, the kernel $K[n](z,w)$ is given by
\begin{align*}
K[n](z,w)=\frac{2^{2n-2}(2n-1)}{\pi}\sum_{\sigma\in \Gamma}
\frac{\sigma'(w)^n}{(z-\sigma (w))^{2n}},\hspace{1cm}z\in \Omega,
w\in\Omega^*.
\end{align*}
The term $\det N_n$ in \eqref{eq61} turned up to be equal to $\det
K[n]K[n]^*$. On the other hand, the Grunsky matrix $A[1]$ can be
considered as the generalization of $K[n]$ to
$\Mob(S^1)\bk\Diff_+(S^1)$ and $n=1$. This gives us some hints of
how to generalize \eqref{eq61} to $\Mob(S^1)\bk\Diff_+(S^1)$ for
arbitrary $n\geq 2$.

In this paper, we start by working with the manifold
$S^1\bk\Diff_+(S^1)$, a fiber space of $\Mob(S^1)\bk\Diff_+(S^1)$
with fiber isomorphic to $\Del^*$.  For any point $\gamma\in
S^1\bk\Diff_+(S^1)$ and any integer $n\geq 1$, we  define bases of
holomorphic $1-n$ differentials and $n$ differentials for
$\Omega=f(\Del)$ and $\Omega^*=g(\Del^*)$, which mimic the
definitions of $\{u[0]_k\}, \{v[0]_k\}$ \eqref{eq65} and $\{U[1]_k,
V[1]_k\}$ \eqref{eq66}. In fact, there are two ways to generalize
$\{U[1]_k\}, \{V[1]_k\}$, one of the them we denote by $\{U[n]_k\}$,
$\{V[n]_k\}$, and the other by $\{p[n]_k\}, \{q[n]_k\}$. The bases
$\{p[n]_k\}$, $\{q[n]_k\}$ are related to the kernel of Bers
integral operator $K[n]:\ov{A_{n,2}(\Omega^*)}\rightarrow
A_{n,2}(\Omega)$ by
\begin{align*}
&K[n](f(z),g(w))f'(z)^n g'(w)^n\\
=&\frac{2^{2n-2}(2n-1)}{\pi}\frac{f'(z)^ng'(w)^n}{(f(z)-g(w))^{2n}}\\
=&\sum_{k=n}^{\infty} p[n]_k(f(z))f'(z)^n \sqrt{\frac{2^{2n-2}}{\pi
(2n-2)!}\frac{(k+n-1)!}{(k-n)!}}w^{-k-n}\\
=&\sum_{k=n}^{\infty} q[n]_k(g(w))g'(w)^n \sqrt{\frac{2^{2n-2}}{\pi
(2n-2)!}\frac{(k+n-1)!}{(k-n)!}}z^{k-n}.
\end{align*}We denote the transition matrices from the basis
$\{U[n]_k\}$ to the basis $\{p[n]_k\}$ and from the basis
$\{V[n]_k\}$ to the basis $\{q[n]_k\}$ by $\mathfrak{P}[n]$ and
$\mathfrak{M}[n]$ respectively. They are upper triangular matrices
with diagonal elements identically equal to $1$. We show that the
matrices $\mathfrak{P}[n]-\Id[n]$ and $\mathfrak{M}[n]-\Id[n]$ are
trace class. We also define the matrices $A[1-n]$, $B[1-n]$,
$C[1-n]$, $D[1-n]$, $A[n]$, $B[n]$, $C[n]$, $D[n]$,
$\mathfrak{A}[n]$, $\mathfrak{B}[n]$, $\mathfrak{C}[n]$,
$\mathfrak{D}[n]$ so that their columns give the coefficients of the
expansions of $u[1-n]_k(f(z))f'(z)^{1-n}$,
$u[1-n]_k(g(z))g'(z)^{1-n}$, $v[1-n](f(z))f'(z)^{1-n}$,
$v[1-n]_k(g(z))g'(z)^{1-n}$, $U[n]_k(f(z))f'(z)^n$,
$U[n]_k(g(z))g'(z)^n$, $V[n]_k(f(z))f'(z)^n$, $V[n]_k(g(z))g'(z)^n$,
$p[n]_k(f(z))f'(z)^n$, $p[n]_k(g(z))g'(z)^n$, $q[n]_k(f(z))f'(z)^n$,
$q[n]_k(g(z))g'(z)^n$ with respect to standard bases of
corresponding differentials on $\Del$ and $\Del^*$. Since
$\mathfrak{A}[n]$ is the matrix of $K[n]$ with respect to the
standard bases of $n$-differentials on $\Del$ and $\Del^*$, by
proving that the Bers integral operator $K[n]$ is a bounded
operator, we can conclude that these matrices all define bounded
operators on some Hilbert spaces.

With the choice of the bases $\{U[n]_k\}, \{V[n]_k\}$ or
$\{p[n]_k\}, \{q[n]_k\}$ of $A_{n,2}(\Del)$ and $A_{n,2}(\Del^*)$,
we define the period matrices $N_n(\Omega)$, $N_n(\Omega^*)$ of
holomorphic $n$ differentials of $\Omega$ and $\Omega^*$ to be the
Gram matrices of these bases with respect to the inner products on
$A_{n,2}(\Omega)$ and $A_{n,2}(\Omega^*)$ respectively. Depending on
the bases we choose, they are equal to $A[n]^T\ov{A[n]}$ and
$D[n]^T\ov{D[n]}$ in the former case, and equal to
$\mathfrak{A}[n]^T\ov{\mathfrak{A}[n]}$ and
$\mathfrak{D}[n]^T\ov{\mathfrak{D}[n]}$ in the later. One of our
main challenge is to show that the function
$\mathfrak{F}_n:\Mob(S^1)\bk \Diff_+(S^1)\rightarrow \R$, where
\begin{align*}
\mathfrak{F}[n] = \log\det N_n(\Omega) =\log\det A[n]A[n]^*
\end{align*}is well-defined. Namely, we need to show that
$A[n]A[n]^*-\Id[n]$ is of trace class. Since
$\mathfrak{A}[n]=A[n]\mathfrak{P}[n]$, and $\mathfrak{P}[n]-\Id[n]$
is a strictly upper triangular trace class operator,
$\mathfrak{F}_n$ can also be interpreted as $\log\det K[n]K[n]^*$.

To show that $A[n]A[n]^*-\Id[n]$ is of trace class, we derive
Grunsky-like equalities which generalize the Grunsky equality
\eqref{eq83} for Grunsky matrices. We first define the $\Z\times \Z$
matrix $\Pi[\gamma;n]$ for any integer $n$ so that its columns are,
up to normalization constants, given by the coefficients of the
expansion of $ \gamma(z)^{k-n}\gamma'(z)^n$, $k\in \Z$ with respect
to $z^{l-n}$. Let
$$\Pi_1[\gamma;n]=(\Pi[\gamma;n])_{l,k\geq
n},\hspace{1cm}\Pi_2[\gamma; n]=(\Pi[\gamma;n]_{-l,k})_{l\geq
1-n,k\geq n}.$$ We show that for $n\geq 1$, there exists matrices
$\mathfrak{S}_j[n], j=1,2,3,4$ such
that\begin{align}\label{eq80}\begin{pmatrix} \Pi_1[\gamma;n ]&
\ov{\Pi_2[\gamma; n]}\\\Pi_2[\gamma;n]&\ov{\Pi_1[\gamma; n]}
\end{pmatrix}\begin{pmatrix}
\Pi_1[\gamma;n ]^*& -\Pi_2[\gamma;
n]^*\\-\ov{\Pi_2[\gamma;n]}^*&\ov{\Pi_1[\gamma; n]}^*
\end{pmatrix}=\begin{pmatrix} \Id[n]+\mathfrak{S}_1[n]&\mathfrak{S}_3[n]\\\mathfrak{S}_2[n]&
\Id[n]+\mathfrak{S}_4[n]\end{pmatrix}\end{align} and
$\mathfrak{S}_1[n]=\ov{\mathfrak{S}_4[n]}$ is trace class. By
showing that\begin{align*} A[\gamma; n]
=&\Pi_1[\gamma^{-1};n]^{-1},\hspace{1.8cm}
B[\gamma;n]=\Pi_2[\gamma^{-1};n]\Pi_1[\gamma^{-1};n]^{-1},\\
C[\gamma;n]=&
\ov{\Pi_2[\gamma;n]}\,\ov{\Pi_1[\gamma;n]}\,^{-1},\hspace{1cm}
D[\gamma;n] =\ov{\Pi_1[\gamma;n]}\,^{-1}\end{align*} for all
integers $n$, we derive from \eqref{eq80} the identity
$$A[\gamma;n] A[\gamma;n]^* =
\Id[n]-A[\gamma;n]\mathfrak{S}_1[\gamma^{-1};n]A[\gamma;n]^*-
C[\gamma;1-n]^T\ov{C[\gamma; 1-n]}.$$ To conclude that $\log\det
A[n]A[n]^*$ is well-defined, we show by using variation techniques
that $C[\gamma;1-n]^T\ov{C[\gamma; 1-n]}$ is trace class.

After the tedious effort spent on proving that the function
$\mathfrak{F}_n$ is well defined, we proceed to compute its
derivative. We show that
\begin{align*}
\pa\mathfrak{F}_n(\mathrm{v}) =\frac{6n^2-6n+1}{12\pi i} \oint_{S^1}
\mathcal{S}(g)(z)\mathrm{v}(z)dz.
\end{align*}From this we conclude the \emph{universal index theorem } on $\Mob(S^1)\bk\Diff_+(S^1)$ :
\begin{align*}
\det N_n =\det K[n]K[n]^*= \exp\left(
-\frac{6n^2-6n+1}{12\pi}S\right)
\end{align*}and\begin{align*}\pa\bar{\pa}\mathfrak{F}_n =-\frac{6n^2-6n+1}{6\pi i}\omega_{WP},\end{align*} which are the generalizations of
\eqref{eq61} and \eqref{eq81} to our homogeneous space
$\Mob(S^1)\bk\Diff_+(S^1)$. It also follows that
\begin{align*}
\det N_n =(\det N_1)^{6n^2-6n+1},
\end{align*}a universal version of Mumford isomorphism \cite{Mum}.
\section{The homogeneous spaces $S^1\bk\Diff_+(S^1)$ and
$\Mob(S^1)\bk\Diff_+(S^1)$} In this section, we briefly recall some
facts we need about the homogeneous spaces $S^1\bk\Diff_+(S^1)$ and
$\Mob(S^1)\bk\Diff_+(S^1)$.

 Let $\Del$ and $\Del^*$ be the unit
circle and its exterior, and let $S^1$ be the unit circle. Denote by
$\Diff_+(S^1)$ the space of orientation preserving diffeomorphisms
on the unit circle $S^1$. Under composition of mappings,
$\Diff_+(S^1)$ is a Fr$\acute{\text{e}}$chet Lie group. We identify
the subgroup of rotations with $S^1$ itself. It defines a left
action on $\Diff_+(S^1)$ and the resulting homogeneous space
$S^1\bk\Diff_+(S^1)$ is a complex K\"ahler manifold, which is an
object of much interest in string theory. According to Kirillov
\cite{K} (see also \cite{Teo}), for every point $\gamma\in
S^1\bk\Diff_+(S^1)$, identified with an element of $\Diff_+(S^1)$
fixing the point 1, there exists a unique conformal welding $\gamma
= g^{-1}\circ f$, where $f$ and $g$ can be extended to
diffeomorphisms on $\hat{\C}$ in such a way that
\begin{itemize}
\item[\textbf{W1}] $f$ is holomorphic on $\Del$ and $g$ is holomorphic on
$\Del^*$.
\item[\textbf{W2}] $f(0)=0, g(\infty)=\infty, f'(0)=1$.
\end{itemize}
We call $(f,g)$ the pair of univalent functions associated to
$\gamma$. The domains $\Omega=f(\Del)$ and $\Omega^*=g(\Del^*)$ are
simply connected domains in $\hat{\C}$ with common boundary
$\mathcal{C}=f(S^1)=g(S^1) $ a $C^{\infty}$ curve.

 Under the
inversion $\mathfrak{I} :S^1\bk\Diff_+(S^1) \rightarrow
S^1\bk\Diff_+(S^1)$, $\mathfrak{I}(\gamma) =\gamma^{-1}$, the pair
of univalent functions $(f[\gamma^{-1}], g[\gamma^{-1}])$ associated
to $\gamma^{-1}$ is related to the pair of univalent functions
$(f[\gamma], g[\gamma])$ associated to $\gamma$ by
\begin{align}\label{eq100}
f[\gamma^{-1}](z) =(\bar{r}\circ\iota\circ g[\gamma]\circ \iota)(z)
=\bar{r}/\ov{g[\gamma](1/\z)},\\
g[\gamma^{-1}](z) =(\bar{r}\circ\iota\circ f[\gamma]\circ \iota)(z)
=\bar{r}/\ov{f[\gamma](1/\z)},\nonumber
\end{align}
where $\iota: \hat{\C}\rightarrow \hat{\C}$ is the inversion
$z\mapsto 1/\z$ on $\hat{\C}$ and $r=g'(\infty)$.

Let $\Mob(S^1)=\PSU(1,1)$ be the group of $\Mobius$ transformations
on the unit circle. The homogeneous space $\Mob(S^1)\bk\Diff_+(S^1)$
is a quotient space of $S^1\bk\Diff_+(S^1)$. We represent a point
$\gamma\in \Mob(S^1)\bk\Diff_+(S^1)$ with an element $\gamma\in
\Diff_+(S^1)$ that fixes $-1, -i, $ and $1$. Fix a point
$\gamma_0\in \Mob(S^1)\bk\Diff_+(S^1)$, the points in the fiber of
$S^1\bk\Diff_+(S^1)\rightarrow \Mob(S^1)\bk\Diff_+(S^1)$ over
$\gamma_0$ are given by $\sigma_w\circ \gamma_0\in
S^1\bk\Diff_+(S^1)$, $w\in \Del^*$, where $\sigma_w$ is the linear
fractional transformation
\begin{align*}
\sigma_w(z)=\frac{1-w}{1-\bar{w}}\frac{1-z\bar{w}}{z-w}.
\end{align*}

The tangent space at the origin of $\Mob(S^1)\bk\Diff_+(S^1)$
consist of smooth vector fields  $$\mathrm{u}(z)
=\sum_{k\in\Z\setminus\{-1,0,1 \}}c_k z^{k+1},
\hspace{1cm}c_{-k}=-\bar{c}_k, \,z\in S^1$$\footnote{Here we
purposely differ from the usual convention by a factor of $iz$.} on
$S^1$. The holomorphic and anti-holomorphic tangent vectors
corresponding to $\mathrm{u}$ are given respectively by
\begin{align}\label{eq85}
\mathrm{v}=\sum_{k=2}^{\infty} c_k z^{k+1}, \hspace{1cm}
\bar{\mathrm{v}}(z)=\ov{\mathrm{v}(z)}=\sum_{k=2}^{\infty}\bar{c}_k
\bar{z}^{k+1}.
\end{align}
The tangent space at any other point $\gamma\in
\Mob(S^1)\bk\Diff_+(S^1)$ can be identified with the tangent space
at the origin via right translation. To be more precise, let
$\gamma_t$ be a smooth curve in $\Mob(S^1)\bk\Diff_+(S^1) $ such
that $\gamma_0=\gamma$. Define $\mathrm{u}_t=
\gamma_t\circ\gamma^{-1}$ and
\begin{align*}
\dot{\mathrm{u}}(z)=\left.\frac{d}{dt}\right|_{t=0}
\mathrm{u}_t(z)=\sum_{k\in \Z} c_k z^{k+1}.
\end{align*}It defines the holomorphic and anti-holomorphic tangent
vectors $\mathrm{v}$ and $\bar{\mathrm{v}}$ with formulas given by
\eqref{eq85}.

The Weil Petersson metric is up to constants, the unique
right-invariant K\"ahler metric on $\Mob(S^1)\bk\Diff_+(S^1)$. At
the tangent space of the origin, it is given by
\begin{align*}
\Vert \mathrm{v}\Vert_{WP}^2 =2\pi\sum_{k=2}^{\infty}(k^3-k)|c_k|^2,
\end{align*}where $\mathrm{v}(z) =\sum_{k=2}^{\infty} c_k z^{k+1}$ is a holomorphic tangent
vector.
\section{Spaces and families of holomorphic differentials}
Given the simply connected domains $\Omega$ and $\Omega^*$, and an
integer $n$, a holomorphic $n$ differential on $\Omega$ (resp.
$\Omega^*$ ) is a holomorphic function $\phi$ on $\Omega$ (resp.
$\Omega^*$) such that
\begin{align*}
\phi(z) = O(1) \hspace{0.3cm} \text{as}\;\; z\rightarrow 0
\hspace{1cm}(\text{resp.}\;\; \phi(z)=
O(z^{-2n})\hspace{0.3cm}\text{as}\;\; z\rightarrow \infty).
\end{align*}

For $E=\Omega$ or $\Omega^*$, we denote by $\mathcal{H}^n(E)$ the
space of holomorphic $n$-differentials on $E$. For $n\geq 1$, we let
$\mathcal{H}^{1-n}_0(E)$ to be the subspace of
$\mathcal{H}^{1-n}(E)$ consisting of all $\phi$ that satisfies
\begin{align*}
\phi(z) =& O(z^{2n-1}) \hspace{0.3cm} \text{as}\;\; z\rightarrow
0,\;\text{if} \;\;E=\Omega \\(\text{resp.}\;\; \phi(z)=&
O(z^{-1})\hspace{0.3cm}\text{as}\;\; z\rightarrow \infty,
\;\text{if}\;\;E=\Omega^*).
\end{align*}

For $n\geq 1$,  the Hilbert space $A_{n,2}(E)$ is defined to be the
subspace of $\mathcal{H}^n(E)$ given by
\begin{align*}
A_{n,2}(E)=\left\{ \phi \;\text{holomorphic on $E$}\; :
\;\Vert\phi\Vert_{n,2}^2=
\iint\limits_{\Omega}|\phi|^2\rho_{E}^{1-n}<\infty\right\}.
\end{align*}Here $\rho_{E}$ is the hyperbolic metric (i.e., metric with constant curvature $-1$) density on
$E$.

When $n\geq 1$,  let
\begin{align}\label{eq1}
\alpha_n =\frac{2^{2n-2}}{(2n-2)!\pi}, \hspace{1cm} \beta_n=
\frac{2^{2n-2}(2n-1)}{\pi}=(2n-1)!\alpha_n,
\end{align}
\begin{align}\label{eq13}
c[n]_k =\begin{cases}\sqrt{\alpha_n}\sqrt{
\frac{(|k|+n-1)!}{(|k|-n)!}},
\hspace{1cm}&\text{if}\;\;|k|\geq n,\\
\sqrt{\alpha_n}, \hspace{1cm}&\text{if}\;\; |k|<n.
\end{cases};
\end{align}
\begin{align}\label{eq14}
c[1-n]_k =
\frac{\alpha_n}{c[n]_k}=\begin{cases}\sqrt{\alpha_n}\sqrt{
\frac{(|k|-n)!}{(|k|+n-1)!}},
\hspace{1cm}&\text{if}\;\;|k|\geq n,\\
\sqrt{\alpha_n}, \hspace{1cm}&\text{if}\;\; |k|<n.
\end{cases}
\end{align}

 It is easy to check that
\begin{align*}
\left\{e^+[n]_k(z) = c[n]_k z^{k-n}\,:\,k\geq n\right\}
\hspace{0.5cm}\text{and} \hspace{0.5cm} \left\{e^-[n]_k(z) = c[n]_k
z^{-k-n}\,:\,k\geq n\right\}\end{align*}are orthonormal bases for
$A_{n,2}(\Del)$ and $A_{n,2}(\Del^*)$ respectively. Since
\begin{align*}
\frac{d^{2n-1}}{dz^{2n-1}}\left( c[1-n]_k z^{\pm k +n-1} \right)=
\begin{cases} (\sgn_n k) c[n]_k z^{\pm k -n}
,\hspace{1cm}&\text{if}\;\; |k|\geq n,\\
0, &\text{if}\;\; |k|<n,\end{cases}
\end{align*}where \begin{align*}
\sgn_n k =\begin{cases} 1, \hspace{0.5cm}&\text{if}\;\; k\geq n,\\
-1, &\text{if}\;\; k<n, \end{cases}
\end{align*}we define
\begin{align*}
&\left\{e^+[1-n]_k(z) = c[1-n]_k z^{k+n-1}\,:\,k\geq
n\right\}\hspace{1cm}\text{and}\\&
  \left\{e^-[1-n]_k(z) = c[1-n]_k
z^{-k+n-1}\,:\,k\geq n\right\}\end{align*}as the corresponding bases
of $\mathcal{H}_0^{1-n}(\Del)$ and $\mathcal{H}_0^{1-n}(\Del^*)$.

 Given
a point $\gamma\in S^1\bk\Diff_+(S^1)$ with associated pair of
univalent functions $(f,g)$ and domains $(\Omega, \Omega^*)$, our
goal is to define some canonical bases of
$\mathcal{H}_0^{1-n}(\Omega)$, $\mathcal{H}_0^{1-n}(\Omega^*)$,
$A_{n,2}(\Omega)$ and $A_{n,2}(\Omega^*)$ for any integer $n\geq 1$.

Given a power series $P(w) =\sum_{k\in \Z} c_k w^k$ and a subset $S$
of integers,  we define $$P(w)_S = \sum_{k\in S} c_kw^k.$$

\subsection{Bases of $\mathcal{H}_0^{1-n}(\Omega)$ and
$\mathcal{H}_0^{1-n}(\Omega^*)$} For $k\geq n$, define the
Faber-type polynomials
\begin{align*}
u[1-n]_k(w) = &c[1-n]_k \left( (g^{-1}(w))^{k+n-1}(g^{-1})'(w)^{1-n}\right)_{\geq 2n-1},\\
v[1-n]_k(w) =& c[1-n]_k\left(
(f^{-1}(w))^{-k+n-1}(f^{-1})'(w)^{1-n}\right)_{\leq -1}.
\end{align*}It is easy to see that $u[1-n]_k$ is a polynomial of
degree $k+n-1$ in $w$ and $v[1-n]_k$ is a polynomial of degree
$k-n+1$ in $w^{-1}$. Moreover, by definition, they belong to
$\mathcal{H}^{1-n}_0(\Omega)$ and $\mathcal{H}^{1-n}_0(\Omega^*)$
respectively. Therefore, $\{ u[1-n]_k\,:\, k\geq n\}$ is a basis for
$\mathcal{H}_0^{1-n}(\Omega)$ and $\{ v[1-n]_k\,:\, k\geq n\}$ is a
basis for $\mathcal{H}^{1-n}_0(\Omega^*)$. For $n=1$, $u[0]_k(w)$
and $v[0]_k(w)$ are up to normalization and the constant terms, the
Faber polynomials (see e.g. \cite{Pom, Duren, Teo2}) of the pair
$(f,g)$. One can check by residue calculus that $u[1-n]_k$ and
$v[1-n]_k$ are encoded in the expansion
\begin{align}\label{eq3}
\frac{g'(z)^n w^{2n-1}}{(g(z)-w)g(z)^{2n-1}} =
&\frac{1}{\alpha_n}\sum_{k=n}^{\infty}u[1-n]_k(w)c[n]_k z^{-k-n},\\
\frac{f'(z)^n}{f(z)-w}=&-\frac{1}{\alpha_n}\sum_{k=n}^{\infty}
v[1-n]_k(w)c[n]_k z^{k-n}.\nonumber
\end{align}
Consider the expansions of $u[1-n]_k\circ f (f')^{1-n},
u[1-n]_k\circ g(g')^{1-n}, v[1-n]_k\circ f (f')^{1-n}, v[1-n]_k\circ
g (g')^{1-n}$ around 0 or $\infty$:
\begin{align}\label{eq38}
u[1-n]_k(f(z))f'(z)^{1-n}=& \sum_{l=n}^{\infty} A[1-n]_{lk} c[1-n]_l
z^{l+n-1},\\
u[1-n]_k(g(z))g'(z)^{1-n}=&c[1-n]_k z^{k+n-1}+\sum_{l=1-n}^{\infty}
B[1-n]_{lk} c[1-n]_l z^{-l+n-1}\nonumber,\\
v[1-n]_k(f(z))f'(z)^{1-n}=&c[1-n]_k z^{-k+n-1}+\sum_{l=1-n}^{\infty}
C[1-n]_{lk}c[1-n]_l z^{l+n-1}\nonumber,\\
v[1-n]_k(g(z))g'(z)^{1-n}=&\sum_{l=n}^{\infty} D[1-n]_{lk}c[1-n]_l
z^{-l+n-1}.\nonumber
\end{align}Then
\begin{align}\label{eq89}
\frac{g'(z)^n f'(w)^{1-n}f(w)^{2n-1}}{(g(z)-f(w))g(z)^{2n-1}} =
&\frac{1}{\alpha_n}\sum_{k=n}^{\infty}\sum_{l=n}^{\infty}A[1-n]_{lk}c[1-n]_lc[n]_k w^{l+n-1}z^{-k-n},\\
\frac{g'(z)^n g'(w)^{1-n}g(w)^{2n-1}}{(g(z)-g(w))g(z)^{2n-1}}
=&\frac{1}{z-w}+
\frac{1}{\alpha_n}\sum_{k=n}^{\infty}\sum_{l=1-n}^{\infty}B[1-n]_{lk} c[1-n]_lc[n]_k w^{-l+n-1}z^{-k-n}\nonumber,\\
\frac{f'(z)^nf'(w)^{1-n}}{f(z)-f(w)}=&\frac{1}{z-w}-\frac{1}{\alpha_n}\sum_{k=n}^{\infty}\sum_{l=1-n}^{\infty}
C[1-n]_{lk} c[1-n]_lc[n]_k w^{l+n-1}z^{k-n}\nonumber,\\
\frac{f'(z)^ng'(w)^{1-n}}{f(z)-g(w)}=&-\frac{1}{\alpha_n}\sum_{k=n}^{\infty}\sum_{l=n}^{\infty}
D[1-n]_{lk} c[1-n]_lc[n]_k w^{-l+n-1}z^{k-n}.\nonumber
\end{align}

We define the matrices $A[1-n], \hat B[1-n], B[1-n], \hat C[1-n],
C[1-n], D[1-n]$ by
\begin{align*}
A[1-n]=&(A[1-n]_{lk})_{l,k\geq n},
\hspace{1.8cm}D[1-n]=(D[1-n]_{lk})_{l,k\geq
n},\\B[1-n]=&(B[1-n]_{lk})_{l,k\geq n}
\hspace{2cm}C[1-n]=(C[1-n]_{lk})_{l,k\geq n},\\
\hat{B}[1-n]=&(B[1-n]_{l,k})_{l\geq 1-n,k\geq
n},\hspace{1cm}\hat{C}[1-n]=(C[1-n]_{l,k})_{l\geq 1-n,k\geq n} .
\end{align*}
Using \eqref{eq100}, it is straightforward to verify from
\eqref{eq89} that under the inversion $\gamma\rightarrow
\gamma^{-1}$ on $S^1\bk\Diff_+(S^1)$,
\begin{align}\label{eq102}
A[\gamma^{-1}; 1-n] = \ov{D[\gamma;1-n]}, \hspace{0.5cm}
\hat{B}[\gamma^{-1};1-n] = \ov{\hat{C}[\gamma;
1-n]},\hspace{0.5cm}B[\gamma^{-1};1-n] = \ov{C[\gamma; 1-n]}
\end{align}for all $\gamma\in S^1\bk\Diff_+(S^1)$.

\subsection{Bases of $A_{n,2}(\Omega)$ and
$A_{n,2}(\Omega^*)$} There are two natural ways to choose bases for
$A_{n,2}(\Omega)$ and $A_{n,2}(\Omega^*)$.

\vspace{0.5cm} \noindent \textbf{First Choice.} For $k\geq n$,
define the Faber-type polynomials
\begin{align*}
U[n]_k(w) =&c[n]_k \left( (g^{-1}(w))^{k-n}(g^{-1})'(w)^n\right)_{\geq 0},\\
V[n]_k(w) =& c[n]_k\left(
(f^{-1}(w))^{-k-n}(f^{-1})'(w)^n\right)_{\leq -2n}.
\end{align*}
It is easy to see that $U[n]_k$ is a polynomial of degree $k-n$ in
$w$, $V[n]_k(w)$ is a polynomial of degree $k+n$ in $w^{-1}$ and
$V[n]_k(w) = O(w^{-2n})$ as $w\rightarrow \infty$. Therefore,
$\{U[n]_k(w)\;:\; k\geq n\}$ and $\{V[n]_k(w)\;:\; k\geq n\}$ are
bases of $A_{n,2}(\Omega)$ and $A_{n,2}(\Omega^*)$ respectively. By
residue calculus, it is easy to verify that the families
$\{U[n]_k(w)\}$ and $\{V[n]_k(w)\}$ can be compactly defined by
\begin{align}\label{eq4}
\frac{g'(z)^{1-n}}{g(z)-w} =&\frac{1}{\alpha_n}\sum_{k=n}^{\infty} U[n]_k(w) c[1-n]_k z^{-k+n-1},\\
\frac{f'(z)^{1-n} f(z)^{2n-1}}{(f(z)-w)
w^{2n-1}}=&-\frac{1}{\alpha_n}\sum_{k=n}^{\infty} V[n]_k(w) c[1-n]_k
z^{k+n-1}.\nonumber
\end{align}From the definition of $U[n]_k$ and $V[n]_k$, it is easy to see that
$U[n]_k \circ f (f')^n$, $U[n]_k\circ g (g')^n$, $V[n]_k\circ f
(f')^n$ and $V[n]_k\circ g (g')^n$ have expansions around $0$ or
$\infty$ of the following form:
\begin{align}\label{eq42}
U[n]_k(f(z)) f'(z)^n = &\sum_{l=n}^{\infty} A[n]_{lk}c[n]_l
z^{l-n},\\
U[n]_k(g(z))g'(z)^n =&c[n]_k
z^{k-n}+\sum_{l=1-n}^{\infty}B[n]_{lk}c[n]_lz^{-l-n}\nonumber,\\
V[n]_k(f(z))f'(z)^n=& c[n]_k z^{-k-n}+\sum_{l=1-n}^{\infty}
C[n]_{lk} c[n]_l z^{l-n}\nonumber,\\
V[n]_k(g(z))g'(z)^n =&\sum_{l=n}^{\infty}D[n]_{lk}c[n]_l
z^{-l-n}.\nonumber
\end{align}It follows that
\begin{align}\label{eq90}
\frac{g'(z)^{1-n}f'(w)^n}{g(z)-f(w)} =&\frac{1}{\alpha_n}\sum_{k=n}^{\infty}
\sum_{l=n}^{\infty}A[n]_{lk}c[n]_l c[1-n]_k w^{l-n}z^{-k+n-1},\\
\frac{g'(z)^{1-n}g'(w)^n}{g(z)-g(w)}
=&\frac{1}{z-w}+\frac{1}{\alpha_n}\sum_{k=n}^{\infty}
\sum_{l=1-n}^{\infty}B[n]_{lk}c[n]_l c[1-n]_k
w^{-l-n}z^{-k+n-1},\nonumber\\\frac{f'(z)^{1-n}f'(w)^n
f(z)^{2n-1}}{(f(z)-f(w))
f(w)^{2n-1}}=&\frac{1}{z-w}-\frac{1}{\alpha_n}\sum_{k=n}^{\infty}\sum_{l=1-n}^{\infty}
C[n]_{lk} c[n]_l
 c[1-n]_k w^{l-n}z^{k+n-1},\nonumber\\
\frac{f'(z)^{1-n}g'(w)^n f(z)^{2n-1}}{(f(z)-g(w))
g(w)^{2n-1}}=&-\frac{1}{\alpha_n}\sum_{k=n}^{\infty}\sum_{l=n}^{\infty}
D[n]_{lk} c[n]_l
 c[1-n]_k w^{-l-n}z^{k+n-1}.\nonumber
\end{align}

We define the matrices $A[n], B[n],\hat{B}[n],C[n], \hat{C}[n],
D[n]$ by
\begin{align*}A[n]=&(A[n]_{lk})_{l,k\geq n},
\hspace{2cm}D[n]=(D[n]_{lk})_{l\geq n,k\geq n},\\
B[n]=&(B[n]_{lk})_{l,k\geq n},\hspace{2cm} C[n]=(C[n]_{lk})_{l\geq
n,k\geq n},\\
\hat{B}[n]=&(B[n]_{lk})_{l\geq 1-n,k\geq
n},\hspace{1.2cm}\hat{C}[n]=(C[n]_{lk})_{l\geq 1-n, k\geq n}.
\end{align*}Under the inversion
$\gamma\rightarrow \gamma^{-1}$ on $S^1\bk\Diff_+(S^1)$,
\begin{align*}
A[\gamma^{-1}; n] = \ov{D[\gamma;n]}, \hspace{0.5cm}B[\gamma^{-1};
n] = \ov{C[\gamma;n]} ,\hspace{0.5cm}\hat{B}[\gamma^{-1};n] =
\ov{\hat{C}[\gamma; n]}
\end{align*}for all $\gamma\in S^1\bk\Diff_+(S^1)$.

Comparing \eqref{eq89} and \eqref{eq90} gives the following
relation.
\begin{align}\label{eq51}
A[1-n]=D[n]^T, \hspace{1cm} D[1-n]=A[n]^T.
\end{align}

\vspace{0.5cm}\noindent \textbf{Second Choice.} Another natural
bases $\{p[n]_k\,:\,k\geq n\}$ and $\{q[n]_k\,:\, k\geq n\}$ of
$A_{n,2}(\Omega)$ and $A_{n,2}(\Omega^*)$ are defined by
\begin{align}\label{eq5} p[n]_k(w)=u[1-n]_k^{(2n-1)}(w),
\hspace{1cm}q[n]_k(w)= -v[1-n]_k^{(2n-1)}(w).
\end{align}It is easy to see that $p[n]_k(w)$ is a polynomial of
degree $k-n$ in $w$, $q[n]_k(w)$ is a polynomial of degree $k+n$ in
$w^{-1}$ and $q[n]_k(w)= O(w^{-2n})$ as $w\rightarrow \infty$.
Therefore, $\{p[n]_k\;|\; k\geq n\}$ and $\{q[n]_k\;|\; k\geq n\}$
indeed form bases of $A_{n,2}(\Omega)$ and $A_{n,2}(\Omega^*)$
respectively. The expansions of $p[n]_k\circ f (f')^n, p[n]_k\circ g
(g')^n, q[n]_k\circ f (f')^n,  q[n]_k\circ g (g')^n$ around $0$ or
$\infty$ have the following form:
\begin{align}\label{eq53}
p[n]_k(f(z))f'(z)^n =&\sum_{l=n}^{\infty} \mathfrak{A}[n]_{lk}c[n]_l
z^{l-n},\\
p[n]_k(g(z))g'(z)^n = & \sum_{l=n}^{k} \mathfrak{P}[n]_{lk}c[n]_l
z^{l-n}+\sum_{l=1-n}^{\infty} \mathfrak{B}[n]_{lk}c[n]_lz^{-l-n},\nonumber\\
q[n]_k(f(z))f'(z)^n=&\sum_{l=n}^{k} \mathfrak{M}[n]_{lk}c[n]_{l}
z^{-l-n}+\sum_{l=1-n}^{\infty} \mathfrak{C}[n]_{lk}c[n]_l z^{l-n},\nonumber\\
q[n]_k(g(z))g'(z)^n =&\sum_{l=n}^{\infty}\mathfrak{D}[n]_{lk}c[n]_l
z^{-l-n}.\nonumber
\end{align}The matrices $\mathfrak{A}[n], \mathfrak{B}[n],
\hat{\mathfrak{B}}[n], \mathfrak{C}[n], \hat{\mathfrak{C}}[n],
\mathfrak{D}[n], \mathfrak{P}[n], \mathfrak{M}[n]$ are defined
respectively by
\begin{align*}\mathfrak{A}[n]=&(\mathfrak{A}[n]_{lk})_{l,k\geq n},
\hspace{2cm}\mathfrak{D}[n]=(\mathfrak{D}[n]_{lk})_{l,k\geq n},\\
\mathfrak{B}[n]=&(\mathfrak{B}[n]_{lk})_{l,k\geq
n},\hspace{2.0cm}\mathfrak{C}[n]=(\mathfrak{C}[n]_{lk})_{l,k\geq
n},\\\hat{\mathfrak{B}}[n]=&(\mathfrak{B}[n]_{lk})_{l\geq 1-n,k\geq
n},\hspace{1.2cm}\hat{\mathfrak{C}}[n]=(\mathfrak{C}[n]_{lk})_{l\geq
1-n,k\geq
n},\\
\mathfrak{P}[n]=&(\mathfrak{P}[n]_{lk})_{l,k\geq n},
\hspace{1.8cm}\mathfrak{M}[n]=(\mathfrak{M}[n]_{lk})_{l,k\geq n}.
\end{align*}Differentiating \eqref{eq3} with respect to $w$ $(2n-1)$ times and using the definition \eqref{eq1}
of $\beta_n$, we have
\begin{align}\label{eq88}
\beta_n\frac{g'(z)^n}{(g(z)-w)^{2n}}=& \sum_{k=n}^{\infty}
u[1-n]_k^{(2n-1)}(w)c[n]_kz^{-k-n}=\sum_{k=n}^{\infty}p[n]_k(w) c[n]_k z^{-k-n},\\
\beta_n\frac{f'(z)^n}{(f(z)-w)^{2n}}=&- \sum_{k=n}^{\infty}
v[1-n]_k^{(2n-1)}(w)c[n]_kz^{-k-n}=\sum_{k=n}^{\infty}q[n]_k(w)c[n]_k
z^{k-n}.\nonumber
\end{align}Therefore,
\begin{align*}
\beta_n\frac{g'(z)^nf'(w)^n}{(g(z)-f(w))^{2n}}=& \sum_{k=n}^{\infty}
\sum_{l=n}^{\infty}\mathfrak{A}[n]_{lk}
c[n]_l c[n]_kw^{l-n}z^{-k-n},\\
=&\sum_{k=n}^{\infty} \sum_{l=n}^{\infty}\mathfrak{D}[n]_{lk} c[n]_l
c[n]_kz^{-l-n}w^{k-n}.
\end{align*}It follows that, \begin{align*}
\mathfrak{A}[n]=\mathfrak{D}[n]^T.
\end{align*}Under the inversion $\gamma\mapsto \gamma^{-1}$, we have
\begin{align*}
\mathfrak{A}[\gamma^{-1}; n] = &\ov{\mathfrak{D}[\gamma;n]},
\hspace{1cm} \mathfrak{P}[\gamma^{-1};n]
= \ov{\mathfrak{M}[\gamma; n]},\\
\mathfrak{B}[\gamma^{-1}; n] =& \ov{\mathfrak{C}[\gamma;n]},
\hspace{1.1cm} \hat{\mathfrak{B}}[\gamma^{-1};n] =
\ov{\hat{\mathfrak{C}}[\gamma; n]}.
\end{align*} for all $\gamma\in S^1\bk\Diff_+(S^1)$.

From the definitions of $p[n]_k$ and $U[n]_k$, we can write $p[n]_k$
as a linear combination of $U[n]_l$ with $n\leq l \leq k$. In fact,
from the expansion of $p[n]_k(g(z))g'(z)^n$ and
$U[n]_k(g(z))g'(z)^n$, it is easy to conclude that
\begin{align}\label{eq6}
p[n]_k = \sum_{l=n}^k \mathfrak{P}[n]_{lk} U[n]_l .
\end{align}Similarly, we have
\begin{align*}
q[n]_k=\sum_{l=n}^{\infty}\mathfrak{M}[n]_{lk}V[n]_l.
\end{align*}Therefore, $\mathfrak{P}[n]$ and $\mathfrak{M}[n]$ are transition
matrices between the bases $\{p[n]_k\}$, $\{U[n]_k\}$ and the bases
$\{q[n]_k\}$ ,$\{V[n]_k\}$ respectively. These give
\begin{align}\label{eq50}
\mathfrak{A}[n]=A[n]\mathfrak{P}[n],
\hspace{1cm}\mathfrak{B}[n]=B[n]\mathfrak{P}[n],\hspace{1cm}\hat{\mathfrak{B}}[n]=\hat{B}[n]\mathfrak{P}[n],\\
\mathfrak{D}[n]=D[n]\mathfrak{M}[n],\hspace{1cm}\mathfrak{C}[n]=C[n]\mathfrak{M}[n],
\hspace{1cm}\hat{\mathfrak{C}}[n]=\hat{C}[n]\mathfrak{M}[n].\nonumber
\end{align}
In case $n=1$, differentiating \eqref{eq4} with respect to $z$ and
compare to \eqref{eq88}, we find that  $p[1]_k = U[1]_k$ and $q[1]_k
= V[1]_k$ for all $k\geq 1$. Therefore,
$\mathfrak{P}[1]=\mathfrak{M}[1]=\Id$ and
\begin{align*}
\mathfrak{A}[1]=A[1], \hspace{1cm}\mathfrak{B}[1]=B[1] ,\hspace{1cm}
\mathfrak{C}[1]=C[1] , \hspace{1cm}\mathfrak{D}[1]=D[1].
\end{align*}

\subsection{A basis of $\mathcal{H}^{1-n}(\Omega)$} For the purpose of proving our main theorem in a later section, we
also need to consider the family of polynomials defined by
\begin{align*}
U[1-n]_k(w) = c[1-n]_k \left(
g^{-1}(w)^{k+n-1}(g^{-1})'(w)\right)_{\geq 0},\hspace{1cm} k\geq
1-n.
\end{align*}Since $U[1-n]_k(w)$ is a polynomial of degree $k+n-1$ in
$w$, $\{ U[1-n]_k\,:\, k\geq 1-n\}$ form a basis of
$\mathcal{H}^{1-n}(\Omega)$. By residue calculus,
\begin{align*}
\frac{g'(z)^n}{g(z)-w} =\frac{1}{\alpha_n} \sum_{k=1-n}^{\infty}
 U[1-n]_k(w)c[n]_kz^{-k-n}.
\end{align*}Let
\begin{align*}
U[1-n]_k(f(z))f'(z)^{1-n}=&\sum_{l=1-n}^{\infty}
\mathbb{A}[1-n]_{lk}c[1-n]_l
z^{l+n-1},\\
U[1-n]_k(g(z))g'(z)^{1-n}=&c[1-n]_k z^{k+n-1}+ \sum_{l=n}^{\infty}
\mathbb{B}[1-n]_{lk}c[1-n]_l z^{-l+n-1},
\end{align*}and define the matrix
\begin{align*}
\mathbb{A}[1-n]=\left(\mathbb{A}[1-n]_{lk}\right)_{l,k\geq 1-n}.
\end{align*}Then,
\begin{align}\label{eq97}
\frac{g'(z)^nf'(w)^{1-n}}{g(z)-f(w)}=&\frac{1}{\alpha_n}
\sum_{k=1-n}^{\infty} \sum_{l=1-n}^{\infty}\mathbb{A}[1-n]_{lk}
c[n]_k c[1-n]_l w^{l+n-1} z^{-k-n}.
\end{align}For $k\geq n$, we also define the holomorphic $n$-differentials
$\mathfrak{U}[n]_k$ by
\begin{align*}
\mathfrak{U}[n]_k(z)
=\frac{1}{\alpha_n}\frac{d^{2n-1}}{dz^{2n-1}}\left(U[1-n]_k(f(z))f'(z)^{1-n}\right)=\frac{1}{\alpha_n}
\sum_{l=n}^{\infty}\mathbb{A}[1-n]_{lk}c[n]_lz^{l-n}.
\end{align*}Since $f$ is a smooth function and $U[1-n]_k(z)$ is a
polynomial, $\mathfrak{U}[n]_k\in A_{n,2}(\Del)$.
\section{Operators on Hilbert spaces}

It is easy to verify that for any $\phi\in A_{n,2}(\Del)$ or
$A_{n,2}(\Del^*)$, we have the following reproducing formula.
\begin{align*}
\phi(z) =\beta_n\iint\limits_{\Del\;\text{or}\,\Del^*}
\frac{\phi(w)\rho(w)^{1-n}}{(1-z\bar{ w})^2}d^2w.
\end{align*}Therefore, the kernel for the identity operator $\id[n]$ on
$A_{n,2}(\Del)$ and $A_{n,2}(\Del^*)$ are given by \begin{align*}
\id[n](z,w)=\frac{\beta_n}{(1-z\bar{w})^{2n}}=\begin{cases}
\sum_{k=n}^{\infty} c[n]_k^2 (z\bar{w})^{k-n},
\hspace{1cm}&\text{if}\;\; z,w \in \Del,\\
\sum_{k=n}^{\infty} c[n]_k^2 (z\bar{w})^{-k-n}, &\text{if}\;\;
z,w\in \Del^*.\end{cases}
\end{align*}

In \cite{Bers}, Bers defined the integral operator $K[n]$ (resp.
$L[n]$) which maps anti-holomorphic $n$-differentials on $\Omega^*$
(resp. $\Omega$) to holomorphic $n$- differentials on $\Omega$
(resp. $\Omega^*$) by
\begin{align*}
(K[n]\bar{\phi})(z) =&\beta_n\iint\limits_{\Omega^*}
\frac{\ov{\phi(w)}\rho(w)^{1-n}}{(z-w)^{2n}}d^2w ,
 \\(L[n]\bar{\phi})(z)
=&\beta_n\iint\limits_{\Omega}
\frac{\ov{\phi(w)}\rho(w)^{1-n}}{(z-w)^{2n}}d^2w.
\end{align*}Here we want to show that $K[n]$ maps $\ov{A_{n,2}(\Omega^*)}$ into
$A_{n,2}(\Omega)$ and it is a bounded operator.
\begin{proposition}
$K[n]$ is a bounded integral operator mapping
$\ov{A_{n,2}(\Omega^*)}$ into $A_{n,2}(\Omega)$.
\end{proposition}\begin{proof}We have to show that there is a constant $M$ such that for all
$\phi\in A_{n,2}(\Omega^*)$
\begin{align*}
\Vert K\bar{\phi}\Vert_{n,2}\leq \Vert \phi\Vert_{n,2}.
\end{align*}The result is well-known for $n=1$ and in this case, $M=1$ will work (see e.g. \cite{TT1}). Let $\eta_1(w)$ (resp. $\eta_2(z)$) to be the distance of $w\in \Omega^*$ (resp. $z\in \Omega$)
 to the boundary of $\Omega^*$ (resp. $\Omega$). Classical
inequality says that (see e.g. \cite{Nag2})$$\frac{1}{4}\leq
\eta_i(z)^2\rho_i(z)\leq 4, \hspace{1cm}i=1,2.$$Therefore, for any
integer $k\geq 1$, \begin{align*}
\frac{\rho(w)^{1-k}}{|z-w|^{2k-2}}\leq
\frac{\rho(w)^{1-k}}{\eta_1(w)^{2k-2}}\leq 4^{k-1}
\end{align*}and
\begin{align*}
\iint\limits_{\Omega^*}\frac{\rho(w)^{1-k}}{|z-w|^{4k}}d^2w\leq &
4^{k-1} \iint\limits_{|z-w|\geq \eta_2(z)}\frac{1}{|z-w|^{2k+2}}d^2w
\\=& 2^{2k-1}\pi \int_{\eta_2(z)}^{\infty}
\frac{rdr}{r^{2k+2}}\\=&\frac{2^{2k-2}\pi}{k}\eta_2(z)^{-2k}\leq
\frac{2^{4k-2}\pi}{k}\rho(z)^k.
\end{align*}This implies that for $n\geq 2$,
\begin{align}\label{eq91}
\Vert
K[n]\bar{\phi}\Vert_{n,2}^2=&\beta_n^2\iint\limits_{\Del}\left|\iint\limits_{\Del^*}
\frac{\ov{\phi(w)}\rho(w)^{1-n}}{(z-w)^{2n}}d^2w\right|^2\rho(z)^{1-n}d^2z\\
\leq
&\beta_n^2\iint\limits_{\Del}\left(\iint\limits_{\Del^*}\frac{\rho(w)^{2-n}}{
|z-w|^{4n-4}}d^2w\right)\left(\iint\limits_{\Del^*}\frac{|\phi(w)|^2\rho(w)^{-n}}
{|z-w|^4}d^2w\right)\rho(z)^{1-n}d^2z\nonumber\\
\leq &
\frac{2^{4n-6}\pi}{n-1}\beta_n^2\iint\limits_{\Del}\iint\limits_{\Del^*}\frac{|\phi(w)|^2\rho(w)^{-n}}
{|z-w|^4}d^2wd^2z.\nonumber
\end{align}By our result in \cite{TT1}, for $w\in \Del^*$,
\begin{align*}
\frac{1}{\pi^2}\iint\limits_{\Del}
\frac{d^2z}{|z-w|^4}=L[1]L[1]^*(w,w) \leq \Id[1](w,w)
=\frac{1}{4\pi}\rho(w).
\end{align*}Therefore, \eqref{eq91} is bounded by
\begin{align*}
\frac{2^{4n-8}\pi^2}{n-1}\beta_n^2\iint\limits_{\Del^*}|\phi(w)|^2
\rho(w)^{1-n}d^2w=\frac{2^{4n-8}\pi^2}{n-1}\beta_n^2\Vert
\phi\Vert_{n,2}^2.
\end{align*}This implies the assertion.
\end{proof} In this proposition, we show that the norm of the operator $K[n]$ is less than or equal
to $(2^{2n-4}\pi /\sqrt{n-1})\beta_n$. In fact, we conjecture that
it is less than or equal to $1$. We gave some justification of this
conjecture in the Appendix.

Under the isomorphism $A_{n, 2}(\Omega) \simeq A_{n,2}(\Del)$ and
$A_{n,2}(\Omega^*)\simeq A_{n,2}(\Del^*)$ induced by $f$ and $g$
respectively, we can consider $K[n]$ (resp. $L[n]$) as an operator
from $\ov{A_{n,2}(\Del^*)}$ (resp. $\ov{A_{n,2}(\Del)}$) to
$A_{n,2}(\Del)$ (resp. $A_{n,2}(\Del^*)$). It is easy to check that
in this perspective, the kernel of $K[n]$ and $L[n]$ are given by
\begin{align*}
\mathfrak{A}[n](z,w) =\frac{f'(z)^n
g'(w)^{n}}{(f(z)-g(w))^{2n}}\hspace{0.5cm}\text{and}\hspace{0.5cm}\mathfrak{D}[n](z,w)
=\frac{g'(z)^n f'(w)^{n}}{(g(z)-f(w))^{2n}}.
\end{align*}Therefore, \begin{lemma}\label{lemma10}The matrices $\mathfrak{A}[n]$ and
$\mathfrak{D}[n]$ define bounded operators on $\ell^2$.\end{lemma}

For any integer $n\geq 1$, we define kernels
\begin{align}\label{eq95}
\mathcal{A}[n](z,w)
=&\sum_{k=n}^{\infty}\sum_{l=n}^{\infty}A[n]_{l,k} c[n]_l c[n]_k
z^{-l-n}
w^{k-n},\\
\mathcal{C}[n](z,w)
=&\sum_{k=n}^{\infty}\sum_{l=n}^{\infty}C[1-n]_{l, k} c[n]_l c[n]_k
w^{l-n} z^{k-n}\nonumber.
\end{align}We are going to show in later sections that $\mathcal{A}[n]$ and $\mathcal{C}[n]$ define bounded operators
on $\ell^2$. Therefore, they can be considered as integral operators
$\mathcal{A}[n]:\ov{A_{n,2}(\Del)}\rightarrow A_{n,2}(\Del^*)$ and
$\mathcal{C}[n] : \ov{A_{n,2}(\Del)}\rightarrow A_{n,2}(\Del)$. From
\eqref{eq90} and \eqref{eq89}, we find that
\begin{align}\label{eq98}
\mathcal{A}[n](z,w)=&-\alpha_n
\frac{d^{2n-1}}{dz^{2n-1}}\left(\frac{g'(z)^{1-n}f'(w)^n}{g(z)-f(w)}\right),\nonumber\\
\mathcal{C}[n](z,w)=&-\alpha_n
\frac{d^{2n-1}}{dw^{2n-1}}\left(\frac{f'(z)^{n}f'(w)^{1-n}}{f(z)-f(w)}-\frac{1}{z-w}\right).
\end{align}Therefore, $\mathcal{A}[n]$ and $
\mathcal{C}[n]^T$ can be considered as the composition of the
integral operators $\mathcal{A}_0[n]$, $\mathcal{C}_0[n]$ mapping
anti-holomorphic $n$ differentials to holomorphic $1-n$
differentials defined by
\begin{align*}
(\mathcal{A}_0[n]\bar{\phi})(z) = &-\alpha_n
\iint\limits_{\Del}\frac{g'(z)^{1-n}f'(w)^n\ov{\phi(w)}\rho(w)^{1-n}}{g(z)-f(w)}d^2w,\\
(\mathcal{C}_0[n]\bar{\phi})(z) = &\alpha_n
\iint\limits_{\Del}\left(\frac{f'(z)^{1-n}f'(w)^n}{f(z)-f(w)}-\frac{1}{z-w}\right)\ov{\phi(w)}\rho(w)^{1-n}d^2w,
\end{align*}with the map $d_n$ of taking $2n-1$ times derivatives of a
$1-n$ differential.

Finally, we also define the operator $\mathcal{D}[n]:
\ov{A_{n,2}(\Del^*)}\rightarrow A_{n,2}(\Del)$ by the kernel
\begin{align}\label{eq199}
\mathcal{D}[n](z,w)
=&\frac{1}{\alpha_n}\sum_{k=n}^{\infty}\sum_{l=n}^{\infty}\mathbb{A}[1-n]_{lk}c[n]_l
c[n]_k
w^{l-n}z^{-k-n}\\
=&\frac{d^{2n-1}}{dw^{2n-1}}\left( \frac{g'(z)^n
f'(w)^{1-n}}{g(z)-f(w)}\right)-\sum_{k=1-n}^{n-1}\mathfrak{U}[n]_k(w)c[n]_k
z^{-k-n}.\nonumber
\end{align}

\section{Period Matrices of holomorphic $n$-differentials}

For a domain $E$, the period matrix  of holomorphic $n$
differentials $N_n(E)$ can be defined after we specify a choice of
basis for $A_{n,2}(E)$. Since there are two natural choices of bases
for $A_{n,2}(\Omega)$ and $A_{n, 2}(\Omega^*)$ respectively, we can
define $N_{n}(\Omega)$ and $N_{n}(\Omega^*)$ in two different ways.

\noindent \textbf{Definition 1.} We define the period matrices
$\mathcal N_n(\Omega)$ and $\mathcal N_n(\Omega^*)$ by the bases $\{
U[n]_k (z)\,:\, k\geq n\}$ and $\{V[n]_k(z)\,:\, k\geq n\}$. More
precisely,
\begin{align*}
\mathcal N_n(\Omega)_{lk} =&\langle U[n]_l,
U[n]_k\rangle_{n,2}=\iint\limits_{\Omega} U[n]_l(w)
\ov{U[n]_k(w)}\rho_{\Omega}(w)^{1-n}d^2w,\\
\mathcal N_n(\Omega^*)_{lk} =&\langle V[n]_l,
V[n]_k\rangle_{n,2}=\iint\limits_{\Omega^*} V[n]_l(w)
\ov{V[n]_k(w)}\rho_{\Omega^*}(w)^{1-n}d^2w.
\end{align*}

\noindent \textbf{Definition 2.} We define the period matrices
$N_n(\Omega)$ and $N_n(\Omega^*)$ by the bases $\{ p[n]_k (z)\,:\,
k\geq n\}$ and $\{q[n]_k(z)\,:\, k\geq n\}$. More precisely,
\begin{align*}
N_n(\Omega)_{lk} =&\langle p[n]_l,
p[n]_k\rangle_{n,2}=\iint\limits_{\Omega} p[n]_l(w)
\ov{p[n]_k(w)}\rho_{\Omega}(w)^{1-n}d^2w,\\
N_n(\Omega^*)_{lk} =&\langle q[n]_l,
q[n]_k\rangle_{n,2}=\iint\limits_{\Omega^*} q[n]_l(w)
\ov{q[n]_k(w)}\rho_{\Omega^*}(w)^{1-n}d^2w.
\end{align*}

Using the fact that $\{ e^+[n]_k\;|\; k\geq n \}$ is an orthonormal
basis for $A_{n,2}(\Del)$, it is easy to compute that
\begin{align*}
\langle U[n]_l, U[n]_k\rangle_{n,2}=&\iint\limits_{\Omega}
U[n]_l(w) \ov{U[n]_k(w)}\rho_{\Omega}(w)^{1-n}d^2w\\
=&\iint\limits_{\Del}
U[n]_l(f(z))f'(z)^n\ov{U[n]_k(f(z))f'(z)^n}\rho_{\Del}(z)^{1-n}d^2z\\
=&\iint\limits_{\Del} \sum_{m=n}^{\infty} A[n]_{ml} c[n]_m z^{m-n}
\sum_{j=n}^{\infty} \ov{A[n]_{jk}c[n]_j
z^{j-n}}\rho_{\Del}(z)^{1-n}d^2z\\
=&\sum_{m=n}^{\infty} A[n]_{ml}\ov{A[n]_{mk}}\\
=&(A[n]^T\ov{A[n]})_{lk}.
\end{align*}This gives
\begin{align*}
\mathcal{N}_n(\Omega) =&
A[n]^T\ov{A[n]}=D[1-n]D[1-n]^*.\end{align*}Similarly, we find
that\begin{align*}
\mathcal{N}_n(\Omega^*) =& D[n]^T\ov{D[n]}=A[1-n]A[1-n]^*,\\
N_n(\Omega) =&
\mathfrak{A}[n]^T\ov{\mathfrak{A}[n]}=\mathfrak{D}[n]\mathfrak{D}[n]^*,\\
N_n(\Omega^*)
=&\mathfrak{D}[n]^T\ov{\mathfrak{D}[n]}=\mathfrak{A}[n]\mathfrak{A}[n]^*.
\end{align*}

\section{The transition matrices $\mathfrak{P}[n]$ and $\mathfrak{M}[n]$}
In this section, we are going to discuss the properties of the
matrices $\mathfrak{P}[n]$ and $\mathfrak{M}[n]$ for any integer
$n\geq 1$.

First, we define the Schwarzian derivative of a diffeomorphism
$T:\hat{\C}\rightarrow\hat{\C}$ by
\begin{align*}
\mathcal{S}(T) =\frac{ T_{zzz}}{T_{z}}
-\frac{3}{2}\left(\frac{T_{zz}}{T_z}\right)^2
=\left(\frac{T_{zz}}{T_z}\right)_z
-\frac{1}{2}\left(\frac{T_{zz}}{T_z}\right)^2.
\end{align*}
Here the subscript $z$ denotes partial derivative with respect to
$z$. We denote $\mathfrak{R}_{\C}[T]$  the ring $\C[\mathcal{S}(T),
\mathcal{S}(T)_z, \mathcal{S}(T)_{zz},\ldots]$ whose elements are
polynomials in $\mathcal{S}(T)^{(n)}, n\geq 0$.\footnote{Here for a
function $F$, $F^{(n)}$ denotes the $n$-partial derivatives with
respect to $z$.} When $T $ is a fractional linear transformation,
$\mathcal{S}(T)=0$ and therefore $\mathfrak{R}_{\C}[T]=\C$.

For $n=1$, we have seen that $\mathfrak{P}[1]=\mathfrak{M}[1]=\Id$.

For $n\geq 2$, we have to apply the following well-known fact:
\begin{lemma}\label{lemma4}
Let $E$ be a domain on $\C$, $h:E\rightarrow \C$ a meromorphic
function on $E$ and $T: \C\rightarrow \C$ a diffeomorphism. For any
integer $n\geq 2$, there exists $\xi[n]_k \in \mathfrak{R}_{\C}[T]$,
$2\leq k\leq 2n-1$ such that
\begin{align*} \left( h\circ T (T_z)^{1-n}\right)^{(2n-1)}
=& h^{(2n-1)}\circ T(T_z)^n - \xi[n]_{2}\left( h\circ T
(T_z)^{1-n}\right)^{(2n-3)}\\&-\xi[n]_3\left( h\circ T
(T_z)^{1-n}\right)^{(2n-4)}-\ldots - \xi[n]_{2n-1}\left( h\circ T
(T_z)^{1-n}\right).
\end{align*}
\end{lemma}
For example, when $n=2$, we have $\xi[2]_2 =2\mathcal{S}(T)$,
$\xi[2]_3=\mathcal{S}(T)_z$, or equivalently
\begin{align*}
\left( \frac{h\circ T}{T_z}\right)_{zzz} = h\circ T
T_{z}^2-2\mathcal{S}(T)\left( \frac{h\circ
T}{T_z}\right)_{z}-\mathcal{S}(T)_z \left( \frac{h\circ
T}{T_z}\right).
\end{align*}
When $n=3$, $\xi[3]_2= 10\mathcal{S}(T)$,
$\xi[3]_3=15\mathcal{S}(T)_z$, $\xi[3]_4=9\mathcal{S}(T)_{zz}
+16\mathcal{S}(T)^2$,
$\xi[3]_5=2\mathcal{S}(T)_{zzz}+16\mathcal{S}(T)\mathcal{S}(T)_z$,
or equivalently,
\begin{align*}
\left( \frac{h\circ T}{T_{zz}}\right)^{(5)} =& h\circ T
T_{z}^3-10\mathcal{S}(T)\left( \frac{h\circ
T}{T_{zz}}\right)_{zzz}-15\mathcal{S}(T)_z \left(
\frac{h\circ T}{T_{zz}}\right)_{zz}\\
&-\left(9\mathcal{S}(T)_{zz}+16\mathcal{S}(T)^2\right)\left(
\frac{h\circ T}{T_{zz}}\right)_z
-\left(2\mathcal{S}(T)_{zzz}+16\mathcal{S}(T)\mathcal{S}(T)_{z}\right)\left(
\frac{h\circ T}{T_{zz}}\right).
\end{align*}

Now, we differentiate the formulas in \eqref{eq4} with respect to
$z$ $(2n-1)$ times:
\begin{align}\label{eq7}
\beta_n\frac{g'(z)^n}{(g(z)-w)^{2n}}=&\sum_{k=n}^{\infty} U[n]_k(w)
c[n]_kz^{-k-n}
\\&+\xi[g;n]_2(z)\sum_{k=n}^{\infty}
U[n]_k(w)c[1-n]_k \frac{(k+n-3)!}{(k-n)!} z^{-k-n+2}\nonumber\\
&-\xi[g;n]_3(z)\sum_{k=n}^{\infty} U[n]_k(w) c[1-n]_k
\frac{(k+n-4)!}{(k-n)!}z^{-k-n+3}\nonumber\\&+\ldots -
\xi[g;n]_{2n-1}(z)\sum_{k=n}^{\infty} U[n]_k(w) c[1-n]_k
z^{-k+n-1}\nonumber.
\end{align}It is easy to check that for $2\leq m \leq 2n-1$,
$(-1)^m\xi[g;n]_m(z)$ has the expansion
\begin{align*}
(-1)^m \xi[g;n]_m(z) =\sum_{k=2}^{\infty} \Xi[g;n;m]_k
c[2]_{k}z^{-k-m}.
\end{align*}For $k<2$, we let $\xi[g;n;m]_k=0$. It follows from \eqref{eq88}, \eqref{eq6} and \eqref{eq7}
that
\begin{align}\label{eq8}
\mathfrak{P}[n]_{lk} = \delta_{lk}
+\frac{c[1-n]_{l}c[2]_{k-l}}{c[n]_k}\sum_{m=2}^{2n-1}\frac{(l+n-m-1)!}{(l-n)!}
\Xi[g;n;m]_{k-l}.
\end{align}

Similarly, we have \begin{align}\label{eq9}
\mathfrak{M}[n]_{lk}=\delta_{lk}
+\frac{c[1-n]_{l}c[2]_{k-l}}{c[n]_k}\sum_{m=2}^{2n-1}\frac{(l+n-1)!}{(l-n+m)!}\Xi[f;n;m]_{k-l},
\end{align}
where
$$\xi[f;n]_{m}(z) =\sum_{k=m}^{\infty}\Xi[f;n;m]_kc[2]_k z^{k-m}.$$
and $\Xi[f;n;m]_k=0$ for $k<m$.

\begin{remark}
In the case $n=2$, we let
$$\mathcal{S}(g)(z) =\sum_{k=2}^{\infty} \mathrm{g}_k c[2]_k
z^{-k-2},\hspace{1cm}\mathcal{S}(f)(z)=\sum_{k=2}^{\infty}\mathrm{f}_k
c[2]_k z^{k-2}.$$ Then
\begin{align*}
\xi[g;2]_2 (z) =& 2\mathcal{S}(g)(z) =\sum_{k=2}^{\infty}
(2\mathrm{g}_k) c[2]_k z^{-k-2}, \\
-\xi[g;2]_3(z) =& -\mathcal{S}(g)'(z)=\sum_{k=2}^{\infty}
((k+2)\mathrm{g}_k) c[2]_k z^{-k-3}\\
\xi[f;2]_2 (z) =& 2\mathcal{S}(f)(z) =\sum_{k=2}^{\infty}
(2\mathrm{f}_k) c[2]_k z^{k-2}, \\
\xi[f;2]_3(z) =& \mathcal{S}(f)'(z)=\sum_{k=3}^{\infty}
((k-2)\mathrm{f}_k) c[2]_k z^{k-3}.
\end{align*}Therefore,
\begin{align*}
\mathfrak{P}[2]_{lk}=&\delta_{lk}
+\frac{c[-1]_{l}c[2]_{k-l}}{c[2]_k}(k+l) \mathrm{g}_{k-l},\\
\mathfrak{M}[2]_{lk}=&\delta_{lk}
+\frac{c[-1]_{l}c[2]_{k-l}}{c[2]_k}(k+l) \mathrm{f}_{k-l}.
\end{align*}\end{remark}

We see from \eqref{eq8} and \eqref{eq9} that the matrices
$\mathfrak{P}[n]$ and $\mathfrak{M}[n]$ can be written  as
\begin{align*}
\mathfrak{P}[n] = \Id + \mathfrak{P}_0[n] , \hspace{1cm}
\mathfrak{M}[n]=\Id +\mathfrak{M}_0[n],
\end{align*}where $\mathfrak{P}_0[n]$ and $\mathfrak{M}_0[n]$ are
strictly upper triangular matrices. We can show more: they are in
fact trace class operators.

\begin{lemma}\label{lemma1}
Let $ K : A_{n,2}(\Del)\rightarrow A_{n,2}(\Del)$ be an operator
with kernel $K(z,w)$. Then
\begin{itemize}
\item[\textbf{I.}]$K$ is a Hilbert-Schmidt operator if and only if\begin{align*}
\iint\limits_{\Del}\iint\limits_{\Del}|K(z,w)|^2\rho(z)^{1-n}\rho(w)^{1-n}d^2z
d^2w<\infty.\end{align*}
\item[\textbf{II.}]
Moreover, if
\begin{align*}
\iint\limits_{\Del}\left(
\iint\limits_{\Del}|K(z,w)|^2\rho(w)^{1-n}d^2w\right)^{1/2}
\rho(z)^{1-(n/2)}d^2z<\infty
\end{align*}or
\begin{align*}
\iint\limits_{\Del}\left(
\iint\limits_{\Del}|K(z,w)|^2\rho(z)^{1-n}d^2z\right)^{1/2}
\rho(w)^{1-(n/2)}d^2w<\infty
\end{align*}
then $K$ is a trace class operator.\end{itemize}
\end{lemma}
\begin{proof}\textbf{I} is well known. If $K$ is Hilbert-Schmidt,
then K is compact. Therefore there exist $\lambda_1\geq \lambda_2
\geq \ldots\geq 0$ and an orthonormal basis $\{ \varphi_j\,|\, j\geq
1\}$ of $A_{n,2}(\Del)$ such that
\begin{align*}
(KK^*)(z,w) = \sum_{j=1}^{\infty} \lambda_j
\varphi_j(z)\ov{\varphi_j(w)} .
\end{align*}Similarly, there exists an orthonormal basis $\{\psi_j\,|\, j\geq 1\}$ of
$A_{n,2}(\Del)$  such that
\begin{align*} (K^*K)(z,w)
=\sum_{j=1}^{\infty}\lambda_j\psi_j(z)\ov{\psi_j(w)}.
\end{align*}$K$ is a trace class operator if and only if
\begin{align*}
\sum_{j=1}^{\infty} \lambda_j^{1/2}<\infty.
\end{align*}By Cauchy-Schwarz inequality,
\begin{align}\label{eq101}
\left(\sum_{j=1}^{\infty}\lambda_j^{1/2}\varphi_j(z)\ov{\varphi_j(z)}\right)^2
\leq \sum_{j=1}^{\infty}\lambda_j
|\varphi_j(z)|^2\sum_{j=1}^{\infty}|\varphi_j(z)|^2.
\end{align}Since
\begin{align*}
\sum_{j=1}^{\infty}\varphi_j(z)\ov{\varphi_j(w)}
\end{align*}is the kernel of the identity operator of
$A_{n,2}(\Del)$, it is equal to $\id[n](z,w)$. Therefore,
\begin{align*}
\sum_{j=1}^{\infty}\varphi_j(z)\ov{\varphi_j(z)}=\frac{\beta_n}{(1-z\bar{z})^{2n}}=\frac{\beta_n}{4^n}
\rho(z)^{n}.
\end{align*}It follows from \eqref{eq101} that
\begin{align*}
\sum_{j=1}^{\infty}\lambda_j^{1/2}\varphi_j(z)\ov{\varphi_j(z)} \leq
2^{-n}\sqrt{\beta_n} \left(\sum_{j=1}^{\infty}\lambda_j
|\varphi_j(z)|^2\right)^{1/2}\rho(z)^{n/2}.
\end{align*}Consequently,
\begin{align*}
\sum_{j=1}^{\infty}\lambda_j^{1/2}=&\iint\limits_{\Del}
\sum_{j=1}^{\infty}\lambda_j^{1/2}\varphi_j(z)\ov{\varphi_j(z)}
\rho(z)^{1-n}d^2z\\
&\leq 2^{-n}\sqrt{\beta_n}
\iint\limits_{\Del}\left(\sum_{j=1}^{\infty}\lambda_j
|\varphi_j(z)|^2\right)^{1/2}\rho(z)^{1-(n/2)}d^2z\\
&=2^{-n}\sqrt{\beta_n}\iint\limits_{\Del}\left(
\iint\limits_{\Del}|K(z,w)|^2\rho(w)^{1-n}d^2w\right)^{1/2}
\rho(z)^{1-(n/2)}d^2z.
\end{align*}Similarly,
\begin{align*}
\sum_{j=1}^{\infty}\lambda_j^{1/2}=&\iint\limits_{\Del}
\sum_{j=1}^{\infty}\lambda_j^{1/2}\psi_j(z)\ov{\psi_j(z)}
\rho(z)^{1-n}d^2z\\
&=2^{-n}\sqrt{\beta_n}\iint\limits_{\Del}\left(
\iint\limits_{\Del}|K(z,w)|^2\rho(z)^{1-n}d^2z\right)^{1/2}
\rho(w)^{1-(n/2)}d^2w.
\end{align*}The assertion \textbf{II} of the lemma follows.
\end{proof}

\begin{proposition}\label{pro1}
For all $2\leq m\leq 2n-1$, the operator $\mathfrak{M}_0[n]$ defines
a trace class operator on $\ell^2$.
\end{proposition}\begin{proof}For $2\leq m\leq 2n-1$, we define the matrices $\mathfrak{M}_m[n]$ by
$$\mathfrak{M}_m[n]_{lk}=\frac{c[1-n]_{l}c[2]_{k-l}}{c[n]_k}\frac{(l+n-1)!}{(l-n+m)!}\Xi[f;n;m]_{k-l},$$
so that
\begin{align*}
\mathfrak{M}_0[n] =\sum_{m=2}^{2n-1}\mathfrak{M}_m[n].
\end{align*}It is sufficient to show that for all
$2\leq m \leq 2n-1$, $\mathfrak{M}_m[n]$ defines a trace class
operator. Denote by $M_m[n]: A_{n,2}(\Del)\rightarrow A_{n,2}(\Del)$
the corresponding operator, i.e. the operator with kernel
\begin{align*}
M_m[n](z,w) =
&\sum_{k=n}^{\infty}\sum_{l=n}^{\infty}\mathfrak{M}_m[n]_{lk}c[n]_l
z^{l-n} c[n]_k \bar{w}^{k-n}\\
=&\sum_{k=n}^{\infty}\sum_{l=n}^{\infty}c[1-n]_{l}c[2]_{k-l}c[n]_l\frac{(l+n-1)!}{(l-n+m)!}\Xi[f;n;m]_{k-l}
z^{l-n}\bar{w}^{k-n}\\
=& \alpha_n
\sum_{k=n+m}^{\infty}\sum_{l=n}^{k-m}c[2]_{k-l}\frac{(l+n-1)!}{(l-n+m)!}\Xi[f;n;m]_{k-l}
z^{l-n}\bar{w}^{k-n}\\
=&\alpha_n\sum_{l=n}^{\infty}\sum_{k=l+m}^{\infty}c[2]_{k-l}\frac{(l+n-1)!}{(l-n+m)!}\Xi[f;n;m]_{k-l}
z^{l-n}\bar{w}^{k-n}\\
=&\alpha_n\sum_{l=n}^{\infty}\sum_{k=m}^{\infty}c[2]_{k}\frac{(l+n-1)!}{(l-n+m)!}\Xi[f;n;m]_{k}
z^{l-n}\bar{w}^{k+l-n}\\
=&\alpha_n\sum_{l=n}^{\infty}\frac{(l+n-1)!}{(l-n+m)!}\xi[f;n]_m(\bar{w})z^{l-n}\bar{w}^{l-n+m}.
\end{align*}

We compute
\begin{align*}
&\iint\limits_{\Del}|M_m[n](z,w)|^2\rho(z)^{1-n} d^2z\\ =
&\alpha_n^2
\sum_{l=n}^{\infty}\left|\frac{(l+n-1)!}{(l-n+m)!}\xi[f;n]_m(\bar{w})\right|^2|w|^{2l-2n+2m}
\frac{1}{c[n]_l^2}\\
=&\alpha_n
\left|\xi[f;n]_m(\bar{w})\right|^2\sum_{l=n}^{\infty}\frac{(l+n-1)!(l-n)!}{(l-n+m)!^2}|w|^{2l-2n+2m}\\
=&\alpha_n
\left|\xi[f;n]_m(\bar{w})\right|^2\sum_{l=n}^{\infty}\frac{(l+n-m-1)!}{(l-n+m)!}|w|^{2l-2n+2m}\frac{(l+n-1)!(l-n)!}{
(l+n-m-1)!(l-n+m)!}.
\end{align*}Since
\begin{align*}
\lim_{l\rightarrow \infty}\frac{(l+n-1)!(l-n)!}{
(l+n-m-1)!(l-n+m)!}=\lim_{l\rightarrow
\infty}\frac{(l+n-1)(l+n-2)\ldots(l+n-m)}{(l-n+m)(l-n+m-1)\ldots(l-n+1)}=1,
\end{align*}There exists a constant $B>0$  such that
\begin{align*}
 \frac{(l+n-1)!(l-n)!}{ (l+n-m-1)!(l-n+m)!} \leq B
\end{align*}
for all $l\geq n$. Therefore,
\begin{align*}
&\iint\limits_{\Del}|M_m[n](z,w)|^2\rho(z)^{1-n} d^2z\\ \leq
&\alpha_n
B\left|\xi[f;n]_m(\bar{w})\right|^2\sum_{l=n-m}^{\infty}\frac{(l+n-m-1)!}{(l-n+m)!}|w|^{2l-2n+2m}\\
=&\alpha_n B (2n-2m-1)!
\frac{\left|\xi[f;n]_m(\bar{w})\right|^2}{(1-|w|^2)^{2n-2m}}\\=&
2^{-2n+2m}\alpha_n B(2n-2m-1)!
\left|\xi[f;n]_m(\bar{w})\right|^2\rho(w)^{n-m}.
\end{align*}Consequently,
\begin{align*}
&\iint\limits_{\Del}\iint\limits_{\Del}|M_m[n](z,w)|^2\rho(z)^{1-n}d^2z\rho(w)^{1-n}d^2w
\\&\leq 2^{-2n+2m}\alpha_n B(2n-2m-1)!
\iint\limits_{\Del}\left|\xi[f;n]_m(\bar{w})\right|^2\rho(w)^{1-m}d^2w.
\end{align*}Since $f$ is smooth, so is $\xi[f;n]_m(\bar{w})$.
Therefore, the last integral is finite and we conclude that $M_m[n]$
is a Hilbert-Schmidt operator. On the other hand, by the same
reasoning, the integral
\begin{align*}
&\iint\limits_{\Del}\left(\iint\limits_{\Del}|M_m[n](z,w)|^2\rho(z)^{1-n}d^2z\right)^{1/2}
\rho(w)^{1-(n/2)}d^2w\\
\leq & 2^{-n+m} \sqrt{\alpha_n
B_2(2n-2m-1)!}\iint\limits_{\Del}\left|\xi[f;n]_m(\bar{w})\right|\rho(w)^{1-(m/2)}d^2w
\end{align*}is also finite. Therefore, $M_m[n]$ is a trace class
operator.

\end{proof}

Using the fact that $\mathfrak{P}[\gamma; n] =
\ov{\mathfrak{M}[\gamma^{-1}; n]}$, we conclude that
\begin{corollary}\label{co3}
For all $n\geq 2$, $\mathfrak{P}_0[ n]$ defines a trace class
operator on $\ell^2$.
\end{corollary}Now it follows from \eqref{eq51}, \eqref{eq50} and
Lemma \ref{lemma10} that
\begin{lemma}\label{lemma12}
For all $n\geq 1$, the matrices $A[\gamma; 1-n]$, $D[\gamma; 1-n]$,
$A[ n]$, $D[n]$ define bounded operators on $\ell^2$.
\end{lemma}
From the definition of $\mathcal{A}[n]$ in \eqref{eq95}, we conclude
that
\begin{corollary}\label{co9}
For all $n\geq 1$, the operator $\mathcal{A}[n]$ is bounded.
\end{corollary}
\section{Identities satisfied by the period matrices}

In this section, we are going to derive some identities satisfied by
the period matrices.

First, we introduce the matrix $\Pi[n]$ generalizing the matrix
$\Pi[0]$ first considered by \cite{Nag92}. Given $\gamma \in
\Diff_+(S^1)$, we can consider the Fourier coefficients
of\footnote{Here $\gamma'= \frac{d\gamma}{dz}$ where
$z=e^{i\theta}$.}
$c[n]_k\gamma(e^{i\theta})^{k-n}\gamma'(e^{i\theta})^n$ for any $k$
and $n$. We define
\begin{align*}
\Pi[\gamma;n]_{lk}=\frac{1}{2\pi}\frac{c[n]_k}{c[n]_l}\int_{0}^{2\pi}\gamma(e^{i\theta})^{k-n}\gamma'(e^{i\theta})^n
e^{-i(l-n)\theta}d\theta,
\end{align*}so that
\begin{align*}
c[n]_k\gamma(e^{i\theta})^{k-n}\gamma'(e^{i\theta})^n=\sum_{l\in\Z}
\Pi[\gamma;n]_{lk} c[n]_l e^{i(l-n)\theta}.
\end{align*}The $\Z\times\Z$ matrix $\Pi[\gamma;n]$ is defined by
$\Pi[\gamma;n]=(\Pi[\gamma;n]_{lk})$. It is easy to see that under
group multiplication,
\begin{align}\label{eq69}\Pi[\gamma_1; n]\Pi[\gamma_2;n]=
\Pi[\gamma_2\circ\gamma_1].\end{align}Therefore
\begin{align}\label{eq12}\Pi[\gamma^{-1};n] =
\Pi[\gamma;n]^{-1}.\end{align} On the other hand, since
$|\gamma(e^{i\theta})|=1$, we have
\begin{align*}
\ov{\gamma(e^{i\theta})} = \gamma(e^{i\theta})^{-1},
\hspace{1cm}\ov{\gamma'(e^{i\theta})} = e^{2i\theta}
\gamma(e^{i\theta})^{-2}\gamma'(e^{i\theta}).
\end{align*}Therefore,
\begin{align*}
c[n]_k\gamma(e^{i\theta})^{-k-n}\gamma'(e^{i\theta})^n =c[n]_k \ov{
\gamma(e^{i\theta})^{k-n} \gamma'(e^{i\theta})^n}e^{-2in\theta}
=\sum_{l\in\Z} \ov{\Pi[\gamma;n]_{lk}} c[n]_l e^{i(-l-n)\theta}.
\end{align*}This implies that
\begin{align}\label{eq11}
\Pi[\gamma;n]_{-l,-k} =\ov{\Pi[\gamma;n]_{l,k}}.
\end{align}On the other hand, since
\begin{align*}
&\frac{1}{2\pi} \int_0^{2\pi}\left( \gamma(e^{i\theta})^{k-n}
\gamma'(e^{i\theta})^n\right)\left( \gamma(e^{i\theta})^{-m+n-1}
\gamma'(e^{i\theta})^{1-n} \right)e^{i\theta}d\theta \\=&
\frac{1}{2\pi}\int_0^{2\pi}
\gamma(e^{i\theta})^{k-m-1}\gamma'(e^{i\theta})e^{i\theta}d\theta\\=&\frac{1}{2\pi
i}\oint_{S^1} z^{k-m-1} dz =\delta_{km},
\end{align*}we have
\begin{align*}
\sum_{l\in\Z} \Pi[\gamma;n]_{l,k}\Pi[\gamma; 1-n]_{-l,-m}
=\delta_{km}.
\end{align*}Together with \eqref{eq11}, we obtain
\begin{align*}
\Pi[\gamma;1-n]^*\Pi[\gamma;n]=\Id.
\end{align*}In view of \eqref{eq12}, we conclude that
\begin{align*}
\Pi[\gamma^{-1}; n] = \Pi[\gamma; 1-n]^*.
\end{align*}We re-collect these identities into the following lemma.

\begin{lemma}\label{lemma5}
For any $\gamma\in \Diff_+(S^1)$ and any integer $n$, we have the
following identities:
\begin{align*}
(i)\;\;&\ov{\Pi[\gamma;n]_{l,k}}=\Pi[\gamma;n]_{-l,-k}.\hspace{6cm}
\\(ii)\;\;& \Pi[\gamma^{-1};n] = \Pi[\gamma;n]^{-1}=\Pi[\gamma;
1-n]^*.\end{align*}
\end{lemma}

Now for $n\geq 1$, let \begin{align*} \Pi_1[\gamma;n] =&
(\Pi[\gamma;n]_{l,k})_{l,k\geq n}, \hspace{2cm}
\hat{\Pi}_2[\gamma;n] = (\Pi[\gamma;n]_{-l,k})_{l \geq 1-n, k\geq
n},\\\hat{\Pi}_3[\gamma;n] =& (\Pi[\gamma;n]_{l,-k})_{l\geq n,k\geq
1-n}, \hspace{1cm} \Pi_4[\gamma;n] = (\Pi[\gamma;n]_{-l,-k})_{l,k
\geq 1-n}\\
 \Pi_2[\gamma;n] =&
(\Pi[\gamma;n]_{-l,k})_{l,k\geq n},
\end{align*}
\begin{align*} \Pi_1[\gamma;1-n] =&
(\Pi[\gamma;1-n]_{l,k})_{l,k\geq n}, \hspace{2cm}
\hat{\Pi}_2[\gamma;1-n] = (\Pi[\gamma;1-n]_{-l,k})_{l \geq 1-n,
k\geq n},\\\hat{\Pi}_3[\gamma;1-n] =&
(\Pi[\gamma;1-n]_{l,-k})_{l\geq n,k\geq 1-n}, \hspace{1cm}
\Pi_4[\gamma;1-n] = (\Pi[\gamma;1-n]_{-l,-k})_{l,k
\geq 1-n}\\
 \Pi_2[\gamma;1-n] =&
(\Pi[\gamma;1-n]_{-l,k})_{l,k\geq n},
\end{align*}
By Lemma \ref{lemma5}, it follows immediately that
\begin{lemma}\label{lemma11}
For any $\gamma\in \Diff_+(S^1)$ and any integer $n$, we have the
following identities:
\begin{align*}
\hspace{1cm}(i)\;\;\Pi_1[\gamma; n]^* =&\Pi_1[\gamma^{-1};1-n],\hspace{8cm}\\
(ii)\;\;\hat{\Pi}_2[\gamma;n]^*=&\hat{\Pi}_3[\gamma^{-1};1-n],\\
(iii)\;\;\Pi_2[\gamma; n]^*=&\ov{\Pi_2[\gamma^{-1}; 1-n]}.
\end{align*}
\end{lemma}

\begin{proposition}\label{pro4}
For every point $\gamma\in S^1\bk\Diff_+(S^1)$, we have the
following relations.

\begin{itemize}\item[\textbf{I.}]
 For any integer $n$,\begin{align*}
A[\gamma;n]=&\Pi_1[\gamma^{-1};n]^{-1},\hspace{2.5cm}D[\gamma;n]=\ov{\Pi_1[\gamma;n]}^{-1}\\
B[\gamma;n]=&\Pi_2[\gamma^{-1};n]\Pi_1[\gamma^{-1};n]^{-1}
,\hspace{1cm}C[\gamma;n]=\ov{\Pi_2[\gamma;n]}\,\ov{\Pi_1[\gamma;n]}^{-1}\\
\hat B[\gamma;n]=&\hat \Pi_2[\gamma^{-1};n]\Pi_1[\gamma^{-1};n]^{-1}
,\hspace{1cm}\hat C[\gamma;n]=\ov{\hat
\Pi_2[\gamma;n]}\,\ov{\Pi_1[\gamma;n]}^{-1}.
\end{align*}\item[\textbf{II.}] For any integer $n\geq 1$,
\begin{align}\label{eq99}
\mathbb{A}[1-n]=\ov{\Pi_4[\gamma^{-1};1-n]}\,^{-1}.
\end{align}\end{itemize}
\end{proposition}
\begin{proof}
For $n\geq 1$, since $f\circ \gamma^{-1}=g$,  we find that
restricted to $S^1$,
\begin{align*}
\left(U[n]_k \circ f (f')^n \right)\circ \gamma^{-1}
((\gamma^{-1})')^n = U[n]_k \circ g (g')^n.
\end{align*}Using the expansion of $U[n]_k\circ f (f')^n$ and
$U[n]_k\circ g (g')^n$ give
\begin{align}\label{eq15}
\Pi_1[\gamma^{-1};n]A[\gamma;n]=\Id, \hspace{1cm}
\Pi_2[\gamma^{-1};n]A[\gamma;n] = B[\gamma;n].
\end{align}The other identities are proved similarly.
\end{proof}
From this, Proposition \ref{pro1}, Corollary \ref{co3} and the fact
that $\mathfrak{P}[n]$ and $\mathfrak{M}[n]$ are invertible, we
conclude that
\begin{lemma}\label{lemma15}For any integer $n\geq 1$,  $A[1-n], A[n], \mathfrak{A}[n], D[1-n], D[n], \mathfrak{D}[n],
\Pi_1[1-n], \Pi_1[n]$ define
 invertible operators on $\ell^2$.\end{lemma}

From part \textbf{II} of Proposition \ref{pro4}, we also have
\begin{corollary}\label{co5}
For any integer $n\geq 1$,  $\mathcal{D}[n]$ is a bounded operator.
\end{corollary}\begin{proof}Let
\begin{align*}
\mathbb{A}_1[1-n] =&\left(\mathbb{A}[1-n]_{l,k}\right)_{l,k\geq n},
\hspace{2.5cm}\mathbb{A}_{2}[1-n]=(\mathbb{A}[1-n]_{lk})_{1-n\leq
l\leq
n-1, k\geq n},\\
\mathbb{A}_3[1-n] =&\left(\mathbb{A}[1-n]_{l,k}\right)_{l\geq
n,1-n\leq k\leq n-1},
\hspace{1cm}\mathbb{A}_{4}[1-n]=(\mathbb{A}[1-n]_{lk})_{1-n\leq l,
k\leq
n-1},\\
\Pi_5[\gamma; 1-n]=&(\Pi[\gamma; 1-n]_{lk})_{1-n\leq l\leq n-1,
k\geq n},\hspace{0.8cm}\Pi_6[\gamma; 1-n]=(\Pi[\gamma;
1-n]_{lk})_{l\geq n,
1-n\leq k\leq n-1},\\
\Pi_7[\gamma; 1-n]=&(\Pi[\gamma; 1-n]_{lk})_{1-n\leq l, k\leq n-1}.
\end{align*}$\mathbb{A}_1[1-n]$ is the kernel of $\mathcal{D}[n]$ with
respect to standard bases. \eqref{eq99} implies that
\begin{align*}
\begin{pmatrix} \mathbb{A}_1[1-n]&\mathbb{A}_3[1-n]\\\mathbb{A}_2[1-n]
&\mathbb{A}_4[1-n]\end{pmatrix}\begin{pmatrix}
\Pi_1[\gamma^{-1};1-n]&\Pi_6[\gamma^{-1};1-n]\\
\Pi_5[\gamma^{-1};1-n]&\Pi_7[\gamma^{-1};1-n]\end{pmatrix}=\Id.
\end{align*}Therefore,
\begin{align*}
\mathbb{A}_1[1-n]\Pi_1[\gamma^{-1}; 1-n]+
\mathbb{A}_3[1-n]\Pi_5[\gamma^{-1}; 1-n]=\Id.
\end{align*}Since $\mathbb{A}_3[1-n]\Pi_5[\gamma^{-1}; 1-n]$ defines a
finite rank operator, it is bounded. Therefore, the bounded-ness of
the operator defined by $\Pi_1[\gamma^{-1}; 1-n]^{-1}$ implies the
bounded-ness of
 $\mathcal{D}[n]$.
\end{proof}

To derive Grunsky-type identities, for $n\geq 1$ we differentiate
both sides of the formula
\begin{align*}
c[1-n]_k\gamma(e^{i\theta})^{k+n-1}\gamma'(e^{i\theta})^{1-n}=\sum_{l\in\Z}
\Pi[\gamma;1-n]_{lk} c[1-n]_l e^{i(l+n-1)\theta}
\end{align*}with respect to $z=e^{i\theta}$ $(2n-1)$ times. Lemma
\ref{lemma4} gives us
\begin{align*}
&(\sgn_n k)c[n]_k \gamma(e^{i\theta})^{k-n}\gamma'(e^{i\theta})^n
\\=&\sum_{|l|\geq n}\Pi[\gamma;1-n]_{lk} (\sgn_n l)
c[n]_le^{i(l-n)\theta}\\
&+\xi[\gamma;n]_2(e^{i\theta}) \sum_{l\in\Z}\Pi[\gamma;1-n]_{lk}
c[1-n]_l
(l+n-1)\ldots (l-n+3) e^{i(l-n+2)\theta}\\
&+\xi[\gamma;n]_3(e^{i\theta}) \sum_{l\in\Z}\Pi[\gamma;1-n]_{lk}
c[1-n]_l
(l+n-1)\ldots(l-n+4) e^{i(l-n+3)\theta}\\
&+\ldots +\xi[\gamma;n]_{2n-1}(e^{i\theta})\sum_{l\in\Z}
\Pi[\gamma;1-n]_{lk} c[1-n]_l e^{i(l+n-1)\theta}.
\end{align*}
 Therefore, for all $|k|\geq n$,
\begin{align}\label{eq19}
&(\sgn_n k) \sum_{l\in\Z}\Pi[\gamma;n]_{lk}c[n]_l e^{i(l-n)\theta}
\\=& \sum_{|l|\geq n}\Pi[\gamma;1-n]_{lk}(\sgn_n l) c[n]_l
e^{i(l-n)\theta}+  \sum_{l\in\Z}
\mathfrak{S}[\gamma;n]_{lj}\Pi[\gamma;1-n]_{jk}(\sgn_n j)c[n]_l
e^{i(l-n)\theta},\nonumber
\end{align}and for $|k|<n$,
\begin{align}\label{eq20} 0= \sum_{|l|\geq
n}\Pi[\gamma;1-n]_{lk}(\sgn_n l) c[n]_l e^{i(l-n)\theta}+
\sum_{l\in\Z} \mathfrak{S}[n]_{lj}\Pi[\gamma;1-n]_{jk}(\sgn_n
j)c[n]_l e^{i(l-n)\theta},
\end{align}
where
\begin{align*}
\mathfrak{S}[\gamma;n]_{lj}=(\sgn_n j)\sum_{m=2}^{2n-1}
\Xi[\gamma;n;m]_{l-j}
\frac{c[2]_{l-j}c[1-n]_{j}}{c[n]_l}(j+n-1)\ldots(j-n+m+1),
\end{align*}and
\begin{align*}
\xi[\gamma;n]_m(e^{i\theta}) =\sum_{k\in\Z} \Xi[\gamma;n;m]_k c[2]_k
e^{i(k-m)\theta}.
\end{align*}\begin{remark}
When $n=2$, $\xi[\gamma;2]_2=2\mathcal{S}(\gamma), $
$\xi[\gamma;2]_3=\mathcal{S}(\gamma)'$. Let
\begin{align*}
\mathcal{S}(\gamma)(e^{i\theta})=\sum_{k\in\Z} \gamma_k c[2]_k
e^{i(k-2)\theta}.
\end{align*}$|\gamma(e^{i\theta})|=1$ implies that
$$\gamma_{-k}=\bar{\gamma}_k.$$A straightforward computation gives
\begin{align*}
\mathfrak{S}[\gamma;2]_{lj} = (\sgn_2 j)
\frac{c[2]_{l-j}c[-1]_j}{c[2]_l}(j+l) \gamma_{l-j}.
\end{align*}
\end{remark}
For any integer $n$, let
\begin{align*}
\tilde{\Pi}[\gamma;n] = \left(
\tilde{\Pi}[\gamma;n]_{lk}\right)_{l,k\in\Z},
\end{align*}where\begin{align*}
\tilde{\Pi}[\gamma;n]_{lk}=&\begin{cases}\Pi[\gamma;n]_{lk},
\hspace{0.2cm}&\text{if}\;\; |k|\geq n,\\
0, &\text{if}\;\; |k|<n.\end{cases}
\end{align*}Also for $n\geq 1$, define
\begin{align*}
J=(J_{lk}), \hspace{0.5cm}\text{where}\hspace{0.5cm}
J_{lk}=\delta_{lk}\sgn_n k,
\end{align*}and
\begin{align*}
\mathfrak{T}[\gamma;n]=I[n] + \mathfrak{S}[\gamma;n]=
\left(I[n]_{lk}+\mathfrak{S}[\gamma;n]_{lk}\right)_{l,k\in
\Z},\end{align*}where
\begin{align*} I[n]_{lk} =\begin{cases}
1,\hspace{0.2cm}&\text{if}\;\;k=l\geq n \;\text{or}\; k=l\leq -n,\\
0, &\text{otherwise}.
\end{cases}
\end{align*}Equations \eqref{eq19} and \eqref{eq20} say that
\begin{align}\label{eq22}
\tilde{\Pi}[\gamma;n] = \mathfrak{T}[\gamma;n]J\Pi[\gamma;
1-n]J=\mathfrak{T}[\gamma;n]J\tilde{\Pi}[\gamma; 1-n]J.
\end{align}Multiplying $J \Pi[\gamma^{-1}; 1-n] J=
J\Pi[\gamma;n]^*J$ on the right of both sides, we find that
\begin{align*}
\tilde{\Pi}[\gamma;n]J\Pi[\gamma;n]^*J = \mathfrak{T}[\gamma;n].
\end{align*}Since the $|k|<n$ columns in $\tilde{\Pi}[\gamma;n]$ are
identically zeros, the $|k|<n$ rows of $J\Pi[\gamma;n]^*J$ do not
contribute anything to the product
$\tilde{\Pi}[\gamma;n]J\Pi[\gamma;n]^*J$. Therefore, we can replace
the $|k|<n$ rows in $J\Pi[\gamma;n]^*J$ by zeros  and
$\tilde{\Pi}[\gamma;n]J\Pi[\gamma;n]^*J=\tilde{\Pi}[\gamma;n]J\tilde{\Pi}[\gamma;n]^*J$.
This implies that
\begin{align}\label{eq21}
\mathfrak{T}[\gamma;n]=\tilde{\Pi}[\gamma;n]J\tilde{\Pi}[\gamma;n]^*J,
\end{align}and therefore
\begin{align}\label{eq30}
\mathfrak{T}[\gamma;n]^*=J\mathfrak{T}[\gamma;n]J.
\end{align}
Let \begin{align*}
\mathfrak{S}_1[\gamma;n]=&(\mathfrak{S}[\gamma;n]_{lk})_{l,k\geq n},
\hspace{1.3cm}\mathfrak{S}_2[\gamma;n]=(\mathfrak{S}[\gamma;n]_{-l,k})_{l,k\geq
n},\\
\mathfrak{S}_3[\gamma;n]=&(\mathfrak{S}[\gamma;n]_{l,-k})_{l,k\geq
n},
\hspace{1cm}\mathfrak{S}_4[\gamma;n]=(\mathfrak{S}[\gamma;n]_{-l,-k})_{l,k\geq
n}.
\end{align*} Equation \eqref{eq30} implies that
\begin{align*}
\mathfrak{S}_1[n]^* =\mathfrak{S}_1[n],
\hspace{1cm}\mathfrak{S}_2[n]^*= -\mathfrak{S}_3[n],
\hspace{1cm}\mathfrak{S}_4[n]^*=\mathfrak{S}_4[n].
\end{align*}
 Now, by deleting the $|l|\leq n-1$
rows and $|k|\leq n-1$ columns of the matrices
$\mathfrak{T}[\gamma;n]$, we find from \eqref{eq21} and \eqref{eq11}
that\begin{lemma} For every integer $n\geq 1$, we have the following
identity.
\begin{align}\label{eq25}\begin{pmatrix} \Id & 0\\0& \Id\end{pmatrix}
+\begin{pmatrix} \mathfrak{S}_1[n] & \mathfrak{S}_3[n]\\
\mathfrak{S}_2[n] & \mathfrak{S}_4[n]
\end{pmatrix}=\begin{pmatrix} \Pi_1[n] & \ov{\Pi_2[n]}\\
\Pi_2[n] & \ov{\Pi_1[n]} \end{pmatrix}\begin{pmatrix} \Pi_1[n]^* & -\Pi_2[n]^*\\
-\ov{\Pi_2[n]}^* & \ov{\Pi_1[n]}^* \end{pmatrix}.
\end{align}\end{lemma}Comparing both sides, we have
\begin{align}\label{eq31}
\Id + \mathfrak{S}_1[n]= &\Pi_1[n]\Pi_1[n]^*
-\ov{\Pi_2[n]}\,\ov{\Pi_2[n]}^*\\
\mathfrak{S}_3[n] = &\ov{\Pi_2[n]}\,\ov{\Pi_1[n]}^*
-\Pi_1[n]\Pi_2[n]^*\nonumber
\end{align}and
\begin{align}\label{eq96}
\mathfrak{S}_4[n]=\ov{\mathfrak{S}_1[n]}, \hspace{1cm}
\mathfrak{S}_3[n]=\ov{\mathfrak{S}_2[n]}.
\end{align}
\begin{remark}
When $n=1$, we have $\mathfrak{S}[1]=0$. Therefore, \eqref{eq25}
says that
\begin{align}\label{eq26}\begin{pmatrix} \Pi_1[1] & \ov{\Pi_2[1]}\\
\Pi_2[1] & \ov{\Pi_1[1]} \end{pmatrix}\begin{pmatrix} \Pi_1[1]^* & -\Pi_2[1]^*\\
-\ov{\Pi_2[1]}^* & \ov{\Pi_1[1]}^* \end{pmatrix}=\Id.
\end{align}On the other hand, by removing the $0$-th row  of
\eqref{eq22}, we find that
\begin{align}\label{eq32}
\begin{pmatrix} \Pi_1[\gamma;1] &\ov{\Pi_2[\gamma;1]}\\
\Pi_2[\gamma;1]& \ov{\Pi_1[\gamma;1]
}\end{pmatrix}=\begin{pmatrix}\Pi_1[\gamma;0] &-\ov{\Pi_2[\gamma;0]}\\
-\Pi_2[\gamma;0]& \ov{\Pi_1[\gamma;0] }\end{pmatrix}.
\end{align}
Together with \eqref{eq26}, this gives
\begin{align*}
 \begin{pmatrix}\Pi_1[\gamma;0] & \ov{\Pi_2[\gamma;0]}\\
\Pi_2[\gamma;0] & \ov{\Pi_1[\gamma;0]} \end{pmatrix}\begin{pmatrix} \Pi_1[\gamma;0]^* & -\Pi_2[\gamma;0]^*\\
-\ov{\Pi_2[\gamma;0]}^* & \ov{\Pi_1[\gamma;0]}^*
\end{pmatrix}=\Id,
\end{align*}a well known identity (see e.g. \cite{Nag92}). From this
equation and Proposition \ref{pro4}, we can derive the Grunsky
equality for the pair $(f,g)$.
\end{remark}

Now we derive the Grunsky-type identities for $(f,g)$ from
\eqref{eq31}. For $n\geq 1$, equation \eqref{eq31} gives
\begin{align*}
\Pi_1[\gamma^{-1};n]^{-1}\bigl(\Pi_1[\gamma^{-1};n]^{-1}\bigr)^* =&
\Id
-\Pi_1[\gamma^{-1};n]^{-1}\mathfrak{S}_1[\gamma^{-1};n]\bigl(\Pi_1[\gamma^{-1};n]^{-1}\bigr)^*
\\&-\Pi_1[\gamma^{-1};n]^{-1}\ov{\Pi_2[\gamma^{-1};n]}\,\ov{\Pi_2[\gamma^{-1};n]}^*
\bigl(\Pi_1[\gamma^{-1};n]^{-1}\bigr)^*,\end{align*}\begin{align*}
&\Pi_1[\gamma^{-1};n]^{-1}
\mathfrak{S}_3[\gamma^{-1};n]\left(\ov{\Pi_1[\gamma^{-1};n]^{-1}}\right)^*\hspace{6cm}\\=&
\Pi_1[\gamma^{-1};n]\ov{\Pi_2[\gamma^{-1};n]}
-\Pi_2[\gamma^{-1};n]^*\left(\ov{\Pi_1[\gamma^{-1};n]^{-1}}\right)^*.
\end{align*}By Lemma \ref{lemma11}, Proposition \ref{pro4}, \eqref{eq102} and \eqref{eq51}, we
have
\begin{align*}&\Pi_1[\gamma^{-1};n]^{-1}=A[\gamma;n]=D[\gamma;
1-n]^T=A[\gamma^{-1}; 1-n]^*=\ov{D[\gamma^{-1};n]},\\
& \ov{\Pi_2[\gamma^{-1};n]}^*\bigl(\Pi_1[\gamma^{-1};n]^{-1}\bigr)^*
=\Pi_2[\gamma; 1-n]\Pi_1[\gamma; 1-n]^{-1} =\ov{C[\gamma;
1-n]}=B[\gamma^{-1}; 1-n].
\end{align*}Therefore, we obtain
\begin{proposition}For any integer $n\geq 1$ and  $\gamma\in S^1\bk\Diff_+(S^1)$, we have the
following identitities.\begin{align}\label{eq49}
&A[\gamma;n]A[\gamma;n]^* = \Id - A[\gamma;n]
\mathfrak{S}_1[\gamma^{-1};n]A[\gamma;n]^* -C[\gamma;
1-n]^T\ov{C[\gamma; 1-n]},\\
&A[\gamma;1-n]^*A[\gamma;1-n] = \Id - A[\gamma;1-n]^*
\mathfrak{S}_1[\gamma;n]A[\gamma;1-n] -B[\gamma;
1-n]^*B[\gamma; 1-n]\nonumber,\\
&D[\gamma;n]D[\gamma;n]^* = \Id - D[\gamma;n]
\mathfrak{S}_4[\gamma;n]D[\gamma;n]^* -B[\gamma;
1-n]^T\ov{B[\gamma; 1-n]},\nonumber\\
&D[\gamma;1-n]^*D[\gamma;1-n] = \Id - D[\gamma;1-n]^*
\mathfrak{S}_4[\gamma^{-1};n]D[\gamma;1-n] -C[\gamma;
1-n]^*C[\gamma;
1-n],\nonumber\\&A[\gamma;n]\mathfrak{S}_3[\gamma^{-1};n]A[\gamma;n]^T
=C [\gamma; 1-n]^T -C [\gamma;
1-n],\nonumber\\&D[\gamma;n]^T\mathfrak{S}_2[\gamma;n]D[\gamma;n] =B
[\gamma; 1-n]^T -B [\gamma; 1-n]\nonumber.
\end{align}\end{proposition}

\section{The functions $\mathfrak{F}_n$ and $\mathfrak{G}_n$ on $S^1\bk \Diff_+(S^1)$.}
In this section, we want to show that we can define real-valued
functions $\mathfrak{F}_n :S^1\bk\Diff_+(S^1)\rightarrow \R$ and
$\mathfrak{G}_n:S^1\bk\Diff_+(S^1)\rightarrow \R$ by $\log\det
A[\gamma;n]A[\gamma;n]^*$ and
$\log\det\mathfrak{A}[n]\mathfrak{A}[n]^*$ respectively.

First, we have the following propositions:
\begin{proposition}\label{pro3}
For any $\gamma\in S^1\bk\Diff_+(S^1)$, the matrix
$\mathfrak{S}_1[\gamma;n]$ defines a trace class operator on
$\ell^2$.
\end{proposition}
\begin{proof}
Since $\mathfrak{S}_1[n]^* = \mathfrak{S}_1[n]$, we can write
\begin{align*}
\mathfrak{S}_1[n]= \mathrm{D}[n] + \mathfrak{L}[n] +
\mathfrak{L}[n]^*,
\end{align*}where $\mathrm{D}[n]$ is a real-valued diagonal matrix and $\mathfrak{L}[n]$ is a strictly lower triangular
matrix.  From the previous section, we find that
\begin{align*}
\mathrm{D}[n]=&\sum_{m=2}^{2n-1} \mathrm{D}_m[n]=\sum_{m=2}^{2n-1}(\mathrm{D}[n]_{lk})_{l,k\geq n},\\
\mathfrak{L}[n]=&\sum_{m=2}^{2n-1} \mathfrak{L}_m[n]
=\sum_{m=2}^{2n-1} (\mathfrak{L}_m[n]_{lk})_{l,k\geq n},
\end{align*}where
\begin{align*}
\mathrm{D}_m[n]_{lk}=&\delta_{l,k} \Xi[\gamma;n;m]_{0}
\frac{c[2]_{0}c[1-n]_{k}}{c[n]_k}\frac{(k+n-1)!}{(k-n+m)!}\\
\mathfrak{L}_m[n]_{lk} =& \begin{cases} \Xi[\gamma;n;m]_{l-k}
\frac{c[2]_{l-k}c[1-n]_{k}}{c[n]_l}\frac{(k+n-1)!}{!(k-n+m)!},
\hspace{1cm}&\text{if}\;\; l>k\\
0, &\text{otherwise}
\end{cases},
\end{align*}and
\begin{align*}
\xi[\gamma;n]_m(e^{i\theta}) =\sum_{k\in\Z} \Xi[\gamma;n;m]_k c[2]_k
e^{i(k-m)\theta}.
\end{align*}It  suffices to show that each of the matrices
$\mathrm{D}_m[n]$, $\mathfrak{L}_m[n]$, $2\leq m\leq 2n-1$ defines a
trace class operator on $\ell^2$. By \eqref{eq13},
\begin{align*}
\sum_{k=n}^{\infty}\frac{c[1-n]_{k}}{c[n]_k}\frac{(k+n-1)!}{(k-n+m+1)!}
=& \sum_{k=n}^{\infty}\frac{1}{(k-n+1)(k-n+2)\ldots (k-n+m)}\\\leq &
\sum_{k=n}^{\infty}\frac{1}{(k-n+1)(k-n+2)} <\infty,
\end{align*} therefore $\mathrm{D}_m[n]$ defines a trace class operator for all
$2\leq m\leq 2n-1$.

To show that $\mathfrak{L}_m[n]$ defines a  trace class operator, we
define
\begin{align*}
\Lambda =(\Lambda_{lk})_{l,k\geq n}, \hspace{0.5cm}\text{where}
\hspace{0.5cm}\Lambda_{l,k}=\delta_{l, k+1}
\end{align*}and write
\begin{align}\label{eq34}
\mathfrak{L}_m[n] =& c[2]_1\xi[\gamma;n;m]_1 \Lambda D_{1,m}[n]+
c[2]_2\xi[\gamma;n;m]_2 \Lambda^2 D_{2,m}[n]\\& +\ldots +
c[2]_{m-1}\xi[\gamma;n;m]_{m-1} \Lambda^{m-1} D_{m-1,m}[n]+
\hat{\mathfrak{L}}_m[n],\nonumber
\end{align}where
\begin{align*}
D_{j,m}[n]=\left( \delta_{l, k}
\frac{c[1-n]_k}{c[n]_{k+j}}\frac{(k+n-1)!}{
(k-n+m)!}\right)_{l,k\geq n},\end{align*}\begin{align*}
\hat{\mathfrak{L}}_m[n]=\bigl(\hat{\mathfrak{L}}_m[n]_{l,k}\bigr),
\hspace{1cm} \hat{\mathfrak{L}}_m[n]_{l,k}=\begin{cases}
\mathfrak{L}_m[n]_{l,k},\hspace{0.5cm}&\text{if}\;\;l\geq k+m\\
0, &\text{otherwise}\end{cases}.
\end{align*}Since $c[n]_{k+j}\geq c[n]_k $ for all $j\geq 1$, we
can show as the case of $\mathrm{D}_m[n]$ that the matrix $D_{j,m}$
is of trace class for all $1\leq j <m\leq 2n-1$. On the other hand,
since the function $\xi[\gamma;n]_m$ is smooth on $S^1$, we can show
as in Proposition \ref{pro1} that $\hat{\mathfrak{L}}_m[n]$ defines
a trace class operator. Since $\Lambda$ defines a bounded operator,
\eqref{eq34} shows that $\mathfrak{L}_m[n]$ defines a trace class
operator. This completes the proof.

\end{proof}

\begin{proposition}\label{Pro4}
For any $\gamma\in S^1\bk\Diff_+(S^1)$ and any integer $n\geq 1$,
the matrix  $C[1-n]$ defines a Hilbert Schmidt operator on $\ell^2$.
\end{proposition}\begin{proof}
For $n=1$, the result is already known (see e.g. \cite{TT1}). Hence
we assume that $n\geq 2$.

From the definition \eqref{eq95}, to show that $C[1-n]$
 defines a Hilbert-Schmidt operator is equivalent to
showing that the operator $\mathcal{C}[n]$ is Hilbert-Schmidt. By
Lemma \ref{lemma1}, we need to prove that
\begin{align*}
\iint\limits_{\Del}\iint\limits_{\Del} \left|\mathcal{C}[n](z,w)
\right|^2 \rho(z)^{1-n}\rho(w)^{1-n}d^2zd^2w <\infty.
\end{align*}
For this purpose, we use variational techniques. From \eqref{eq98},
we have
\begin{align}\label{eq41}\mathcal{C}[n](z,w)=&-\alpha_n
\frac{d^{2n-1}}{dw^{2n-1}}\left(\frac{f'(z)^{n}f'(w)^{1-n}}{f(z)-f(w)}-\frac{1}{z-w}\right).\end{align}Given
a point $\gamma\in S^1\bk\Diff_+(S^1)$, we can joint it to the
origin $\id\in S^1\bk\Diff_+(S^1)$ by a smooth curve $\gamma_t$,
$t\in [0,1]$ such that $\gamma_0=\id$ and $\gamma_1=\gamma$. We have
the following well-known fact \cite{K}:
\begin{lemma}
Given a smooth curve $\gamma_t\in S^1\bk \Diff_+(S^1)$ and its
associated pair $(f^t, g_t)$, we have
\begin{align*}
\left.\frac{df^{t+s}}{ds}\right|_{s=0}(z) =\frac{1}{2\pi
i}\oint_{S^1}\frac{ f^t(z)^2 g_t'(\zeta)^2 u_t(\zeta)
}{g_t(\zeta)^2(g_t(\zeta)-f^t(z))}d\zeta,
\end{align*}where
\begin{align*}
u_t(\zeta) =\left. \frac{d \gamma_{t+s}}{ds}\right|_{s=0}\circ
\gamma_t^{-1}(\zeta)
\end{align*}is a smooth function on $S^1$.
\end{lemma}
For $\zeta\in S^1$, let
\begin{align*}
u_t(\zeta)=\sum_{k\in \Z} c_{k}(t) \zeta^{k+1}
\end{align*}and define
\begin{align*}
v_t(\zeta) =\sum_{k=1}^{\infty} c_{k}(t)
\zeta^{k+1},\hspace{1cm}\zeta\in S^1.
\end{align*}It is easy to see that
\begin{align*}
\frac{1}{2\pi i}\oint_{S^1}\frac{ f^t(z)^2 g_t'(\zeta)^2 u_t(\zeta)
}{g_t(\zeta)^2(g_t(\zeta)-f^t(z))}d\zeta=\frac{1}{2\pi
i}\oint_{S^1}\frac{ f^t(z)^2 g_t'(\zeta)^2 v_t(\zeta)
}{g_t(\zeta)^2(g_t(\zeta)-f^t(z))}d\zeta.
\end{align*}


Now a straight-forward computation gives
\begin{lemma}\begin{align}\label{eq40}
\left.\frac{d}{ds}\right|_{s=0}\frac{(f^{t+s})'(z)^n
(f^{t+s})'(w)^{1-n}}{f^{t+s}(z)-f^{t+s}(w)}=& \frac{n}{2\pi i}
\oint_{S^1}\frac{v_t(\zeta) g_t'(\zeta)^2 (f^t)'(z)^n
(f^t)'(w)^{1-n}}{(g_t(\zeta)-f^t(w))(g_t(\zeta)-f^t(z))^2}d\zeta
\\&-\frac{1-n}{2\pi i}\oint_{S^1}\frac{v_t(\zeta)
g_t'(\zeta)^2 (f^t)'(z)^n
(f^t)'(w)^{1-n}}{(g_t(\zeta)-f^t(w))^2(g_t(\zeta)-f^t(z))}d\zeta.\nonumber
\end{align}
\end{lemma}
Let
\begin{align*}
\frac{g_t'(\zeta)^n (f^t)'(w)^{1-n}}{g_t(\zeta)-f^t(w)}
=&\frac{1}{\alpha_n}\sum_{k=1-n}^{\infty}\sum_{l=1-n}^{\infty}
\mathbb{A}_t[1-n]_{lk} c[n]_k
\zeta^{-k-n}c[1-n]_l w^{l+n-1},\\
\frac{g_t'(\zeta)^{n+1} (f^t)'(w)^{1-n}}{(g_t(\zeta)-f^t(w))^2}
=&\frac{1}{\alpha_n}\sum_{k=1-n}^{\infty}\sum_{l=1-n}^{\infty}
\mathbb{H}_t[n]_{lk} c[n+1]_k \zeta^{-k-n-1}c[1-n]_l w^{l+n-1},
\end{align*}and define
\begin{align}\label{eq46}
\hat{\mathcal{D}}_t[n](\zeta,
w)=&\frac{1}{\alpha_n}\sum_{k=1-n}^{\infty}\sum_{l=n}^{\infty}
\mathbb{A}_t[1-n]_{lk}
c[n]_k \zeta^{-k-n}c[n]_l w^{l-n},\\
 H_t[n](\zeta, w) =&\frac{1}{\alpha_n}\sum_{k=1-n}^{\infty}\sum_{l=n}^{\infty}
\mathbb{H}[n]_t[n]_{lk} c[n+1]_k \zeta^{-k-n-1}c[n]_l
w^{l-n}.\nonumber
\end{align}

 Comparing \eqref{eq40} with \eqref{eq41},
we find that
\begin{corollary}\label{co1}
\begin{align*}
\left.\frac{d}{ds}\right|_{s=0}
\mathcal{C}[n]_{t+s}(z,w)=&\alpha_n\Biggl(-\frac{n}{2\pi i
}\oint_{S^1}\frac{v_t(\zeta) g_t'(\zeta)^{2-n} (f^t)'(z)^n
}{(g_t(\zeta)-f^t(z))^2}\hat{\mathcal{D}}[n]_t(\zeta,w)d\zeta
\\&+\frac{1-n}{2\pi i}\oint_{S^1}\frac{v_t(\zeta)
g_t'(\zeta)^{1-n} (f^t)'(z)^n
}{g_t(\zeta)-f^t(z)}H_t(\zeta,w)d\zeta\Biggr).
\end{align*}
\end{corollary}
Now we prove the following.
\begin{lemma}
\begin{align*}
\iint\limits_{\Del}\iint\limits_{\Del}\left|\left.\frac{d}{ds}\right|_{s=0}
\mathcal{C}[n]_{t+s}(z,w)\right|^2\rho(z)^{1-n}d^2z\rho(w)^{1-n}d^2w
\leq \mathfrak{K}_t \sum_{k=1}^{\infty}k^{2n+2} |c_k(t)|^2
\end{align*}for some constant $\mathfrak{K}_t$ that depends continuously on $t$.
\end{lemma}\begin{proof}

By \eqref{eq199}, we can write
\begin{align*}
\hat{\mathcal{D}}_t(\zeta,w)=R_t[n](\zeta,w)
+\mathcal{D}_t[n](\zeta,w),
\end{align*}where
\begin{align*}
R_t[n](\zeta,w) =&\sum_{k=1-n}^{n-1}\mathfrak{U}_t[n]_k(w)c[n]_k
\zeta^{-k-n}.
\end{align*}Since
\begin{align*}
\frac{g_t'(\zeta)^{n+1}
(f^t)'(w)^{1-n}}{(g_t(\zeta)-f^t(w))^2}=-\frac{\pa}{\pa\zeta}\left(\frac{g_t'(\zeta)^n
(f^t)'(w)^{1-n}}{g_t(\zeta)-f^t(w)}\right)+\frac{g_t^{\prime\prime}(\zeta)}{g_t'(\zeta)}\frac{g_t'(\zeta)^n
(f^t)'(w)^{1-n}}{g_t(\zeta)-f^t(w)},
\end{align*}we have from the definition \eqref{eq46},
\begin{align*}
H_t(\zeta, w) = &-\frac{\pa
\hat{\mathcal{D}}_t[n](\zeta,w)}{\pa\zeta}+\frac{g_t^{\prime\prime}(\zeta)}{g_t'(\zeta)}\hat{\mathcal{D}_t}[n]_t(\zeta,w)\\
=&-\frac{\pa
R_t[n](\zeta,w)}{\pa\zeta}+\frac{g_t^{\prime\prime}(\zeta)}{g_t'(\zeta)}R_t[n](\zeta,w)-\frac{\pa
\mathcal{D}_t[n](\zeta,w)}{\pa\zeta}+\frac{g_t^{\prime\prime}(\zeta)}{g_t'(\zeta)}\mathcal{D}_t[n](\zeta,w).
\end{align*}By standard reproducing formulas, we have
\begin{align*}
\mathcal{D}_t[n](\zeta,w) =&\beta_n\iint\limits_{\Del^*}
\frac{\mathcal{D}_t[n](v,w)\rho(v)^{1-n}}{(1-\zeta \bar{v})^{2n}}d^2v,\\
\frac{g_t^{\prime\prime}(\zeta)}{g_t'(\zeta)}=&\frac{1}{\pi}\iint\limits_{\Del^*}
\frac{g_t^{\prime\prime}(\eta)}{g_t'(\eta)}\frac{1}{(1-\bar{\eta}\zeta)^2}d^2\eta.
\end{align*}

 On the other hand, by \eqref{eq90} and the definition of $\mathcal{A}[n]$ in \eqref{eq95}, we
can verify directly that
\begin{align*}
\frac{g_t'(\zeta)^{1-n} (f^t)'(z)^n}{g_t(\zeta) - f^t(z)}
=&\frac{1}{\alpha_n}\sum_{k=n}^{\infty}\sum_{l=n}^{\infty}A_t[n]_{lk}c[1-n]_k
\zeta^{-k+n-1}c[n]_l z^{l-n}\\
=&\iint\limits_{\Del^*}\frac{\mathcal{A}_t[n](u,z)\rho(u)^{1-n}}{\bar{u}^{2n-1}(\bar{u}\zeta-1)}d^2u.
\end{align*}
Differentiating with respect to $\zeta$, we have
\begin{align*}
\frac{g_t'(\zeta)^{2-n}f_t'(z)^n}{(g_t(\zeta)-f^t(z))^2}=\iint\limits_{\Del}
\frac{
\mathcal{A}_t[n](u,z)\rho(u)^{1-n}}{(\bar{u}\zeta-1)^2\bar{u}^{2n-2}}d^2u+
\frac{g_t^{\prime\prime}(\zeta)}{g_t'(\zeta)}\iint\limits_{\Del}
\frac{
\mathcal{A}[n]_t(u,z)\rho(u)^{1-n}}{(\bar{u}\zeta-1)\bar{u}^{2n-1}}d^2u.
\end{align*}

From these, we can write
\begin{align*}
&\frac{1}{2\pi i}\oint_{S^1}\frac{v_t(\zeta) g_t'(\zeta)^{2-n}
(f^t)'(z)^n }{(g_t(\zeta)-f^t(z))^2}\hat{\mathcal{D}}_t(\zeta,w)d\zeta\\
=& \frac{1}{2\pi i}\oint_{S^1}\iint\limits_{\Del^*} \frac{v_t(\zeta)
\mathcal{A}_t[n](u,z)R_t[n](\zeta,w)}{(\bar{u}\zeta-1)^2\bar{u}^{2n-2}}\rho(u)^{1-n}d^2ud\zeta\\&+\frac{1}{2\pi^2
i}\oint_{S^1}\iint\limits_{\Del^*}\iint\limits_{\Del^*}
\frac{v_t(\zeta)
\mathcal{A}_t[n](u,z)R_t[n](\zeta,w)}{\bar{u}^{2n-1}(\bar{u}\zeta-1)(1-\bar{\eta}\zeta)^2}\frac{g_t^{\prime\prime}(\eta)}{g_t'(\eta)}\rho(u)^{1-n}d^2\eta
d^2ud\zeta\\
&+ \frac{\beta_n}{2\pi
i}\oint_{S^1}\iint\limits_{\Del^*}\iint\limits_{\Del^*}\frac{v_t(\zeta)
\mathcal{A}_t[n](u,z)\mathcal{D}_t[n](v,w)}{\bar{u}^{2n-2}(\bar{u}\zeta-1)^2(1-\bar{v}\zeta)^{2n}}
\rho(v)^{1-n}\rho(u)^{1-n} d^2vd^2ud\zeta\\&+\frac{\beta_n} {2\pi^2
i}\oint_{S^1}\iint\limits_{\Del^*}\iint\limits_{\Del^*}\iint\limits_{\Del^*}
\frac{v_t(\zeta)
\mathcal{A}_t[n](u,z)\mathcal{D}_t[n](v,w)}{\bar{u}^{2n-1}(\bar{u}\zeta-1)(1-\bar{v}\zeta)^{2n}(1-\bar{\eta}\zeta)^2}
\frac{g_t^{\prime\prime}(\eta)}{g_t'(\eta)}\\
&\hspace{8cm}\rho(u)^{1-n} \rho(v)^{1-n}d^2\eta d^2vd^2ud\zeta,
\end{align*}
and
\begin{align*}
&\frac{1}{2\pi i}\oint_{S^1}\frac{v_t(\zeta) g_t'(\zeta)^{1-n}
(f^t)'(z)^n }{g_t(\zeta)-f^t(z)}H_t(\zeta,w)d\zeta\\
=&-\frac{1}{2\pi
i}\oint_{S^1}\iint\limits_{\Del^*}\frac{v_t(\zeta)\mathcal{A}_t[n](u,z)}{\bar{u}^{2n-1}(\bar{u}\zeta-1)}\frac{\pa
R_t[n](\zeta,w)}{\pa\zeta}\rho(u)^{1-n}d^2u d^2\zeta\\
&+\frac{1}{2\pi^2
i}\oint_{S^1}\iint\limits_{\Del^*}\iint\limits_{\Del^*}
\frac{v_t(\zeta)\mathcal{A}_t[n](u,z)}{\bar{u}^{2n-1}(\bar{u}\zeta-1)(1-\zeta\bar{\eta})^2}
\frac{g_t^{\prime\prime}(\eta)}{g_t'(\eta)}R_t[n](\zeta,w)\rho(u)^{1-n}d^2\eta d^2u d\zeta\\
&-\frac{2n\beta_n}{2\pi i}\oint_{S^1}\iint\limits_{\Del^*}
\iint\limits_{\Del^*}\frac{v_t(\zeta)\mathcal{A}_t[n](u,z)\mathcal{D}_t[n](v,w)\bar{v}}
{\bar{u}^{2n-1}(\bar{u}\zeta-1)(1-\bar{v}\zeta)^{2n+1}}
\rho(u)^{1-n}\rho(v)^{1-n}d^2u d^2vd\zeta\\
&+\frac{\beta_n}{2\pi^2
i}\oint_{S^1}\iint\limits_{\Del^*}\iint\limits_{\Del^*}\iint\limits_{\Del^*}
\frac{v_t(\zeta)\mathcal{A}_t[n](u,z)\mathcal{D}_t[n](v,w)}
{\bar{u}^{2n-1}(\bar{u}\zeta-1)(1-\bar{v}\zeta)^{2n}(1-\zeta\bar{\eta})^2}
\frac{g_t^{\prime\prime}(\eta)}{g_t'(\eta)}\\
&\hspace{8cm}\rho(u)^{1-n}\rho(v)^{1-n}d^2\eta d^2ud^2v d\zeta.
\end{align*}

Since from Corollary \ref{co9} and Corollary \ref{co5}, both
$\mathcal{A}_t[n]$ and $\mathcal{D}_t[n]$ are bounded operators, we
have
\begin{align*}
&\iint\limits_{\Del}\iint\limits_{\Del}\left|\frac{\beta_n}{2\pi
i}\oint_{S^1}\iint\limits_{\Del^*}\iint\limits_{\Del^*}
\frac{v_t(\zeta)
\mathcal{A}_t[n](u,z)\mathcal{D}_t[n](v,w)}{\bar{u}^{2n-2}(\bar{u}\zeta-1)^2(1-\bar{v}\zeta)^{2n}}
\rho(v)^{1-n}\rho(u)^{1-n}d^2vd^2ud\zeta\right|^2\\
&\hspace{6cm}\times
\rho(z)^{1-n}d^2z\rho(w)^{1-n}d^2w\\
\leq &\kappa_{1,t}\kappa_{2,t}
\iint\limits_{\Del^*}\iint\limits_{\Del^*}\left|\frac{1}{2\pi i}
\oint_{S^1}\frac{v_t(\zeta)\beta_n
}{\bar{u}^{2n-2}(\bar{u}\zeta-1)^2(1-\bar{v}\zeta)^{2n}}d\zeta\right|^2
\rho(v)^{1-n}\rho(u)^{1-n}d^2vd^2u \\=&\kappa_{1,t}\kappa_{2,t}
\iint\limits_{\Del^*}\iint\limits_{\Del^*}\left|
\sum_{k=n}^{\infty}\sum_{m=n}^{\infty} c_{k+m}(t)c[n]_m^2(k-n+1)
\bar{v}^{-m-n}\bar{u}^{-k-n} \right|^2 \rho(v)^{1-n}\rho(u)^{1-n}d^2vd^2u\\
=&\kappa_{1,t}\kappa_{2,t} \sum_{k=n}^{\infty}\sum_{m=n}^{\infty}
|c_{k+m}(t)|^2
\frac{c[n]_m^2}{c[n]_k^2}(k-n+1)^2\\
=&\kappa_{1,t}\kappa_{2,t} \sum_{k=2n}^{\infty}
\left(\sum_{j=n}^{k-n}\frac{c[n]_{k-j}^2}{c[n]_j^2}(j-n+1)^2\right)|c_k(t)|^2\\=&\kappa_{1,t}\kappa_{2,t}
\sum_{k=2n}^{\infty}
\left(\sum_{j=n}^{k-n}\frac{(k-j-n+1)\ldots(k-j+n-1)}{(j-n+1)\ldots(j+n-1)}(j-n+1)^2\right)|c_k(t)|^2\\
\leq &\kappa_{1,t}\kappa_{2,t} \sum_{k=2n}^{\infty}k^{2n}|c_k(t)|^2,
\end{align*}where
\begin{align*}
\kappa_{1,t}=\Vert
\mathcal{A}_t[n]\Vert^2_{\infty},\hspace{1cm}\kappa_{2,t}=\Vert
\mathcal{D}_t[n]\Vert_{\infty}^2
\end{align*}are the squares of the sup-norms of the operators $\mathcal{A}_t[n]$ and $\mathcal{D}_t[n]$ respectively.
Similarly, we have
\begin{align}\label{eq45}
&\iint\limits_{\Del}\iint\limits_{\Del}\Biggl|\frac{\beta_n}{2\pi^2
i}\oint_{S^1}\iint\limits_{\Del^*}
\iint\limits_{\Del^*}\iint\limits_{\Del^*}\frac{v_t(\zeta)
\mathcal{A}_t[n](u,z)\mathcal{D}_t[n](v,w)}{\bar{u}^{2n-1}(\bar{u}\zeta-1)
(1-\bar{v}\zeta)^{2n}(1-\bar{\eta}\zeta)^2}\frac{g_t^{\prime\prime}(\eta)}{g_t'(\eta)}
\\&\hspace{3cm}\times\rho(v)^{1-n}\rho(u)^{1-n}d^2\eta d^2vd^2ud\zeta\Biggr|^2
\rho(z)^{1-n}d^2z\rho(w)^{1-n}d^2w\nonumber\\\nonumber\leq
&\kappa_{1,t}\kappa_{2,t}
\iint\limits_{\Del^*}\iint\limits_{\Del^*}\left| \frac{1}{2\pi^2
i}\oint_{S^1}\iint\limits_{\Del^*}\frac{v_t(\zeta)\beta_n
}{\bar{u}^{2n-1}(\bar{u}\zeta-1)(1-\bar{v}\zeta)^{2n}(1-\bar{\eta}\zeta)^2}\frac{g_t^{\prime\prime}(\eta)}{g_t'(\eta)}d^2\eta
d\zeta\right|^2 \\
&\hspace{9cm}\rho(v)^{1-n}\rho(u)^{1-n}d^2vd^2u .\nonumber
\end{align} From our
result in \cite{TT1}, for $\gamma_t\in S^1\bk\Diff_+(S^1)$,
\begin{align*} \varkappa_t
=\iint\limits_{\Del^*}\left|\frac{g_t^{\prime\prime}(\eta)}{g_t'(\eta)}\right|^2d^2\eta<\infty.
\end{align*}
It follows from Cauchy-Schwarz inequality that \eqref{eq45} is
bounded by
\begin{align*}
\leq &\kappa_{1,t}\kappa_{2,t}\varkappa_t
\iint\limits_{\Del^*}\iint\limits_{\Del^*}\iint\limits_{\Del^*}\left|\frac{1}{2\pi^2
i}\oint_{S^1}\frac{v_t(\zeta)\beta_n
}{\bar{u}^{2n-1}(\bar{u}\zeta-1)(1-\bar{v}\zeta)^{2n}(1-\bar{\eta}\zeta)^2}d\zeta\right|^2\\
&\hspace{8cm}d^2\eta
\rho(v)^{1-n}\rho(u)^{1-n}d^2vd^2u\\
=&\kappa_{1,t}\kappa_{2,t}\varkappa_t\sum_{k=1}^{\infty}
\sum_{m=n}^{\infty}\sum_{l=n}^{\infty}|c_{k+m+l}(t)|^2
\frac{c[1]_k^2c[n]_m^2}{c[n]_l^2}\\
\leq
&\kappa_{1,t}\kappa_{2,t}\varkappa_t\pi^{-1}\sum_{k=2n+1}^{\infty}
k^{2n+1}|c_k(t)|^2.
\end{align*}
Now we consider the term
\begin{align*}
&\iint\limits_{\Del}\iint\limits_{\Del}\left|\frac{2n\beta_n}{2\pi
i}\oint_{S^1}\iint\limits_{\Del^*}
\iint\limits_{\Del^*}\frac{v_t(\zeta)\mathcal{A}_t[n](u,z)\mathcal{D}_t[n](v,w)\bar{v}}
{\bar{u}^{2n-1}(\bar{u}\zeta-1)(1-\bar{v}\zeta)^{2n+1}}
\rho(u)^{1-n}\rho(v)^{1-n}d^2ud^2v d\zeta\right|^2\\
&\hspace{8cm}\times\rho(z)^{1-n}d^2z\rho(w)^{1-n}d^2w.\end{align*}
Using the same reasoning, it is bounded by \begin{align*}
&\kappa_{1,t}\kappa_{2,t}\iint\limits_{\Del^*}\iint\limits_{\Del^*}\left|\frac{1}{2\pi
i}\oint_{S^1}
\frac{2nv_t(\zeta)\bar{v}\beta_n}{\bar{u}^{2n-1}(\bar{u}\zeta-1)(1-\bar{v}\zeta)^{2n+1}}
 d\zeta\right|^2\rho(u)^{1-n}d^2u\rho(v)^{1-n}d^2v\\
\leq
&\kappa_{1,t}\kappa_{2,t}\sum_{m=n}^{\infty}\sum_{l=n}^{\infty}|c_{m+l}(t)|^2
\frac{c[n]_m^2}{c[n]_l^2}(m+n)^2\\
\leq
&\kappa_{1,t}\kappa_{2,t}\sum_{k=2n}^{\infty}k^{2n+2}|c_k(t)|^2.
\end{align*}

For the terms containing $R_t[n](\zeta, w)$, we have first
\begin{align*}
&\iint\limits_{\Del}\iint\limits_{\Del}\left|\frac{1}{2\pi
i}\oint_{S^1}\iint\limits_{\Del^*} \frac{v_t(\zeta)
\mathcal{A}_t[n](u,z)R_t[n](\zeta,w)}{\bar{u}^{2n-2}(\bar{u}\zeta-1)^2}\rho(u)^{1-n}d^2ud\zeta\right|^2\\
&\hspace{8cm}
\rho(z)^{1-n}\rho(w)^{1-n}d^2zd^2w\\
\leq
&\kappa_{1,t}\iint\limits_{\Del}\iint\limits_{\Del^*}\left|\frac{1}{2\pi
i}\oint_{S^1} \frac{v_t(\zeta)
R_t[n](\zeta,w)}{\bar{u}^{2n-2}(\bar{u}\zeta-1)^2}d\zeta\right|^2
\rho(u)^{1-n}\rho(w)^{1-n}d^2u d^2w\\
\leq & (2n-1)\kappa_{1,t}
\sum_{l=1-n}^{n-1}\iint\limits_{\Del}\iint\limits_{\Del^*}\left|\frac{1}{2\pi
i}\oint_{S^1} \frac{v_t(\zeta) \mathfrak{U}_t[n]_l(w)
c[n]_l\zeta^{-l-n}}{\bar{u}^{2n-2}(\bar{u}\zeta-1)^2}d\zeta\right|^2
\\&\hspace{8cm}\rho(u)^{1-n}\rho(w)^{1-n}d^2u d^2w\\
 =&(2n-1)\kappa_{1,t}\alpha_n^{-1}
\sum_{l=1-n}^{n-1}c[n]_l^2\Vert
\mathfrak{U}_t[n]_l\Vert_{n,2}^2\left(\sum_{k=n}^{\infty}|c_{k+l}(t)|^2
\frac{(k-n+1)^2}{(k-n+1)\ldots (k+n-1)}\right)\\
\leq &(2n-1)\kappa_{1,t}\sum_{l=1-n}^{n-1}\Vert
\mathfrak{U}_t[n]_l\Vert_{n,2}^2\sum_{k=1}^{\infty}|c_k(t)|^2.
\end{align*}

Similarly,
\begin{align*}
&\iint\limits_{\Del}\iint\limits_{\Del}\Biggl|\frac{1}{2\pi^2
i}\oint_{S^1}\iint\limits_{\Del^*}\iint\limits_{\Del^*}
\frac{v_t(\zeta)
\mathcal{A}_t[n](u,z)R_t[n](\zeta,w)}{\bar{u}^{2n-1}(\bar{u}\zeta-1)(1-\bar{\eta}\zeta)^2}\frac{g_t^{\prime\prime}(\eta)}{g_t'(\eta)}
d^2\eta\rho(u)^{1-n}d^2ud\zeta\Biggr|^2\\&\hspace{8cm}
\rho(z)^{1-n}\rho(w)^{1-n}d^2zd^2w\\
\leq & (2n-1)\kappa_{1,t}\varkappa_t
\sum_{l=1-n}^{n-1}\iint\limits_{\Del}\iint\limits_{\Del^*}\iint\limits_{\Del^*}\left|\frac{1}{2\pi^2
i}\oint_{S^1} \frac{v_t(\zeta) \mathfrak{U}_t[n]_l(w)
c[n]_l\zeta^{-l-n}}{\bar{u}^{2n-1}(\bar{u}\zeta-1)(1-\bar{\eta}\zeta)^2}d\zeta\right|^2\\&\hspace{8cm}\times
d^2\eta\rho(u)^{1-n}\rho(w)^{1-n}d^2u d^2w\\
 =&(2n-1)\frac{\kappa_{1,t}\varkappa_t}{\alpha_n\pi}
\sum_{l=1-n}^{n-1}c[n]_l^2\Vert
\mathfrak{U}_t[n]_l\Vert_{n,2}^2\left(\sum_{k=n}^{\infty}\sum_{m=1}^{\infty}|c_{k+m+l}(t)|^2
\frac{m}{(k-n+1)\ldots (k+n-1)}\right) \\\leq
&(2n-1)\frac{\kappa_{1,t}\varkappa_t}{\pi} \sum_{l=1-n}^{n-1}\Vert
\mathfrak{U}_t[n]_l\Vert_{n,2}^2\sum_{k=2}^{\infty}k|c_k(t)|^2.\end{align*}

Finally,
\begin{align*}
&\iint\limits_{\Del}\iint\limits_{\Del}\left|\frac{1}{2\pi
i}\oint_{S^1}
\iint\limits_{\Del^*}\frac{v_t(\zeta)\mathcal{A}_t[n](u,z)}{\bar{u}^{2n-1}(\bar{u}\zeta-1)}\frac{\pa
R_t[n](\zeta,w)}{\pa\zeta}\rho(u)^{1-n}d^2u d\zeta\right|^2\rho(z)^{1-n}\rho(w)^{1-n}d^2zd^2w\\
&\leq
(2n-1)\kappa_{1,t}\sum_{l=1-n}^{n-1}\iint\limits_{\Del}\iint\limits_{\Del^*}\left|\frac{1}{2\pi
i}\oint_{S^1} \frac{v_t(\zeta)
c[n]_l(l+n)\mathfrak{U}_t[n]_l(w)\zeta^{-l-n-1}}{\bar{u}^{2n-1}(\bar{u}\zeta-1)}d\zeta\right|^2\\
&\hspace{8cm}\times \rho(u)^{1-n}\rho(w)^{1-n}d^2u d^2w\\
&\leq (2n-1)\kappa_{1,t}\sum_{l=1-n}^{n-1}(l+n)^2\Vert
\mathfrak{U}_t[n]_l\Vert_{n,2}^2\sum_{k=1}^{\infty}|c_{k}(t)|^2.
\end{align*}Putting everything together, we find that
\begin{align*}
\iint\limits_{\Del}\iint\limits_{\Del}\left|\left.\frac{d}{ds}\right|_{s=0}
\mathcal{C}[n]_{t+s}(z,w)\right|^2\rho(z)^{1-n}d^2z\rho(w)^{1-n}d^2w
\leq \mathfrak{K}_t \sum_{k=1}^{\infty}k^{2n+2} |c_k(t)|^2,
\end{align*}where
\begin{align*}
\mathfrak{K}_t
=&\alpha_n^2\kappa_{1,t}\kappa_{2,t}\left(n^2+(1-n)^2+(2n-1)^2\frac{\varkappa_t}{\pi}\right)\\
+&(2n-1)\kappa_{1,t}\alpha_n^2\sum_{l=1-n}^{n-1}\left(n^2+(1-n)^2(l+n)^2+(2n-1)^2
\frac{\varkappa_t}{\pi}\right)\Vert \mathfrak{U}_t[n]_l\Vert_{n,2}^2
\end{align*}depends continuously on $t$, since
each of the terms $\kappa_{1,t}$, $\kappa_{2,t}$, $\varkappa_t$,
$\Vert \mathfrak{U}_t[n]_l\Vert_{n,2}^2$ depends smoothly on $t$.
\end{proof}
 Now we come back to show that $\mathcal{C}[n]$ is Hilbert Schmidt.
 Since $\mathcal{C}[n]=0$ when $\gamma=\id$, we have
\begin{align*}
&\iint\limits_{\Del}\iint\limits_{\Del} \left|\mathcal{C}[n](z,w)
\right|^2
\rho(z)^{1-n}\rho(w)^{1-n}d^2zd^2w  \\
=&\iint\limits_{\Del}\iint\limits_{\Del} \left|\int_0^1 \frac{d
\mathcal{C}[n]_t(z,w)}{dt}dt \right|^2 \rho(z)^{1-n}\rho(w)^{1-n}d^2zd^2w \\
\leq & \int_0^1\iint\limits_{\Del}\iint\limits_{\Del} \left| \frac{d
\mathcal{C}[n]_t(z,w)}{dt} \right|^2 \rho(z)^{1-n}\rho(w)^{1-n}d^2zd^2w dt\\
\leq &\int_{0}^1 \mathfrak{K}_t\sum_{k=1}^{\infty}k^{2n+2}
|c_k(t)|^2 dt<\infty,
\end{align*}since both $ \mathfrak{K}_t$ and $\sum_{k=1}^{\infty}k^{2n+2}
|c_k(t)|^2$ depend continuously on $t$.

\end{proof}
It follows from Proposition \ref{pro4} and Lemma \ref{lemma15} that
\begin{corollary}
For any $\gamma\in S^1\bk\Diff_+(S^1)$ and any $n\in\Z$, the
matrices $\Pi_2[\gamma;n]$, $B[\gamma;n]$, $C[\gamma;n]$ define
Hilbert-Schmidt operators on $\ell^2$.
\end{corollary}

Since we have shown in Proposition \ref{pro3} and Proposition
\ref{Pro4} that $\mathfrak{S}_1[n]$ and $C[\gamma;
1-n]^T\ov{C[\gamma; 1-n]}$ are trace class operators, from the
identity \eqref{eq49}
\begin{align*}
A[\gamma;n]A[\gamma;n]^* = \Id - A[\gamma;n]
\mathfrak{S}_1[\gamma^{-1};n]A[\gamma;n]^* -C[\gamma;
1-n]^T\ov{C[\gamma; 1-n]},
\end{align*}we conclude that the Fredholm determinant of
$A[\gamma;n]A[\gamma;n]^*$ is well defined for all $n\geq 1$. Now
since $A[\gamma;1-n]=A[\gamma^{-1}; n]^*$,  the Fredholm determinant
of $A[\gamma;n]A[\gamma;n]^*$ is well defined for all integers $n$.
On the other hand, since $\mathfrak{A}[\gamma;
n]=A[\gamma;n]\mathfrak{P}[\gamma;n]$, Corollary \ref{co3} implies
that the Fredholm determinant of
$\mathfrak{A}[\gamma;n]\mathfrak{A}[\gamma;n]^*$ is also well
defined for all $n\geq 1$. Hence we can make the following
definition:

\begin{definition}
For any $n\in\Z$, we define the real--valued function
$\mathfrak{F}_n : S^1\bk\Diff_+(S^1)\rightarrow \R$ by
\begin{align*}
\mathfrak{F}_n =\log \det A[\gamma;n]A[\gamma;n]^*.
\end{align*}
For any integer $n\geq 1$, we define the real--valued function
$\mathfrak{G}_n : S^1\bk\Diff_+(S^1)\rightarrow \R$ by
\begin{align*}
\mathfrak{G}_n =\log \det
\mathfrak{A}[\gamma;n]\mathfrak{A}[\gamma;n]^*=\log\det
(K[n]K[n]^*)=\log\det N_n(\Omega^*).
\end{align*}\end{definition}
Notice that for $n\geq 1$, $$\mathfrak{F}_{1-n}(\gamma) =\log\det
\mathcal{N}_n(\Omega^*).$$

Some properties of the functions $\mathfrak{F}_n$ and
$\mathfrak{G}_n$ are listed below.
\begin{proposition}We have the followings:

\begin{itemize}\item[ \textbf{A.}]\;\; For any integer $n$, $\mathfrak{F}_n(\gamma)
=\mathfrak{F}_{1-n}(\gamma^{-1})$.

\item[ \textbf{B.}] For any integer $n\geq 1$, $\mathfrak{G}_n$ is
invariant with respect to the inversion $\mathfrak{I}$ on
$S^1\bk\Diff_+(S^1)$, i.e., $\mathfrak{G}_n(\gamma)
=\mathfrak{G}_{n}(\gamma^{-1})$.

\item[ \textbf{C.}] For any integer $n\geq 1$,
$\mathfrak{F}_n=\mathfrak{G}_n=\mathfrak{F}_{1-n}$.

\item[ \textbf{D.}]For any integer $n$,
$$\hspace{1cm}\mathfrak{F}_n(\gamma)=-\log\det
\Pi_1[\gamma;n]\Pi_1[\gamma;n]^*=-\log\det
\Pi_1[\gamma;1-n]\Pi_1[\gamma;1-n]^*.$$

\item[ \textbf{E.}] Both the functions $\mathfrak{F}_n$ and
$\mathfrak{G}_n$ are  constant on each fiber of the manifold
$S^1\bk\Diff_+(S^1)$ over $\Mob(S^1)\bk\Diff_+(S^1)$. Hence they
descend to well-defined functions on $\Mob(S^1)\bk\Diff_+(S^1)$,
which we denote by the same symbols.

\end{itemize}
\end{proposition}

\begin{proof}Notice that for any operator $K$ where $KK^*-\id$ is a trace class operator, $\det
KK^*=\det K^*K$. \textbf{A.} follows from the identity
\begin{align*}
A[\gamma;1-n]=A[\gamma^{-1};n]^*
\end{align*}and \textbf{B.} follows from
\begin{align*}
\mathfrak{A}[\gamma^{-1};n]=\ov{\mathfrak{D}[\gamma;n]}=
\mathfrak{A}[\gamma;n]^*.
\end{align*}

It is well known in Fredholm theory that if both $K_1$ and $K_2$ are
operators such that $K_1-\id$ and $K_2-\id$ are trace class
operators, then $\det K_1K_2=\det K_1\det K_2=\det K_2K_1$. On the
other hand, since $\mathfrak{P}[\gamma;n]$ is an upper triangular
matrix with diagonal elements all equal to $1$,  $\det
\mathfrak{P}[\gamma;n]=\det\mathfrak{P}[\gamma;n]^*=1$. It follows
that
\begin{align*}
\det \mathfrak{A}[\gamma;n]\mathfrak{A}[\gamma;n]^* =&\det
\mathfrak{A}[\gamma;n]^*\mathfrak{A}[\gamma;n] \\=&\det\left(
\mathfrak{P}[\gamma;n]^*A[\gamma;n]^*A[\gamma;n]\mathfrak{P}[\gamma;n]\right)\\
=&\det\left(
\mathfrak{P}[\gamma;n]^*\right)\det\left(A[\gamma;n]^*A[\gamma;n]\right)
\det\left(\mathfrak{P}[\gamma;n]\right)\\
=&\det\left(A[\gamma;n]A[\gamma;n]^*\right),
\end{align*}
i.e. $\mathfrak{G}_n =\mathfrak{F}_n$.  We can then conclude the
other equality in \textbf{C.} by \textbf{A.} and \textbf{B.}.

Now, since $A[\gamma^{-1};n]=\Pi_1[\gamma;n]^{-1}$, by \textbf{B.}
and \textbf{C.},
\begin{align*}
\mathfrak{F}_n(\gamma) =\mathfrak{F}_n(\gamma^{-1})=
\log\det\left(A[\gamma^{-1};n]A[\gamma^{-1};n]^*\right)=-\log\det\left(\Pi_1[\gamma;n]\Pi_1[\gamma;n]^*\right).
\end{align*}Together with \textbf{C.} prove \textbf{D.}.

Finally, when $\sigma\in \PSU(1,1)$ is a linear fractional
transformation, $\mathcal{S}(\sigma)=0$. Therefore,
$\mathfrak{S}[\sigma;n]=0$. On the other hand, since $\sigma$
extends to a holomorphic function on $\Del$, we have
\begin{align}\label{eq54}
\hat{\Pi}_2[\sigma;n]=0\hspace{1cm}\text{for all\;\;$n\geq 1$. }
\end{align}It follows from \eqref{eq31} that
\begin{align}\label{eq55}
\Pi_1[\sigma;n]\Pi_1[\sigma;n]^* =\Id.
\end{align}By \eqref{eq69}
\begin{align*}\Pi[\sigma\circ\gamma_0]=\Pi[\gamma_0;n]\Pi[\sigma;n]\end{align*} and
\eqref{eq54}, we conclude that
\begin{align}\label{eq56}
\Pi_1[\sigma\circ\gamma_0] =\Pi_1[\gamma_0;n]\Pi_1[\sigma;n].
\end{align}It follows from \textbf{D.}, \eqref{eq55} and \eqref{eq56} that
\begin{align*}
\mathfrak{F}_n(\sigma\circ \gamma_0)
=&-\log\det\left(\Pi_1[\sigma\circ\gamma_0;n]\Pi_1[\sigma\circ\gamma_0;n]^*\right)\\=&
-\log\det\left(\Pi_1[\gamma_0;n]\Pi_1[\gamma_0;n]^*\right)=\mathfrak{F}_n(\gamma_0).
\end{align*}Therefore, $\mathfrak{F}_n$ is invariant on each fiber
of $S^1\bk\Diff_+(S^1)\rightarrow\Mob(S^1)\bk\Diff_+(S^1)$. By
\textbf{C.}, the same holds for $\mathfrak{G}_n$.

\end{proof}

\section{First and Second derivatives of the functions $\mathfrak{F}_n$ and $\mathfrak{G}_n$ on
$\Mob(S^1)\bk\Diff_+(S^1)$}

In this section, we compute the first derivatives  of the function
$\mathfrak{F}_n=\mathfrak{G}_n$.

We begin by an interesting lemma.\begin{lemma}Let $E$ be a domain
and  $h: E\rightarrow\hat{\C}$ a univalent function. Then we have
the following formula:
\begin{align*}
\lim_{w\rightarrow z} \left(\left( n\frac{\pa}{\pa z}
+(n-1)\frac{\pa}{\pa
w}\right)\left[\frac{h'(z)^{1-n}h'(w)^n}{h(z)-h(w)}-\frac{1}{z-w}\right]\right)=-\frac{6n^2-6n+1}{6}
\mathcal{S}(h)(z).
\end{align*}

\end{lemma}The proof is just some calculus. However, this formula is essential
in our theorem below. I am grateful to A. Mcintyre who pointed out
this formula to me a few years ago.
\begin{theorem}
Let $\gamma\in \Mob(S^1)\bk \Diff_+(S^1)$ and $\dot{\mathrm{u}} =
(\mathrm{v} , \bar{\mathrm{v}})$ a tangent vector at $\gamma$. The
first derivative of the function $\mathfrak{F}_n$ is given by
\begin{align*}
\pa\mathfrak{F}_n (\mathrm{v}) = &\frac{6n^2-6n+1}{12\pi
i}\oint_{S^1}\mathcal{S}(g)(z) \mathrm{v}(z)dz\\
\bar{\pa}\mathfrak{F}_n (\bar{\mathrm{v}}) =
&-\frac{6n^2-6n+1}{12\pi i}\oint_{S^1}\ov{\mathcal{S}(g)(z)}\,\bar{
\mathrm{v}}(z)d\z.
\end{align*}
\end{theorem}
\begin{proof}
We use the definition
\begin{align*}
\mathfrak{F}_n (\gamma) = -\log\Pi_1[\gamma; 1-n]\Pi_1[\gamma;
1-n]^*
\end{align*}for the function $\mathfrak{F}_n$.

Let $\gamma_t$ be a smooth curve in $\Mob(S^1)\bk\Diff_+(S^1)$ which
defines the tangent vector $\dot{\mathrm{u}} = (\mathrm{v} ,
\bar{\mathrm{v}})$ at $\gamma_0=\gamma$. It is a standard fact that
\begin{align}\label{eq701}
&\left.\frac{d}{dt}\right|_{t=0}\log\det(\Pi_1
[\gamma_t;1-n]\Pi_1[\gamma_t;1-n]^*)\\
=&\Tr \left((\Pi_1
[\gamma;1-n]\Pi_1[\gamma;1-n]^*)^{-1}\left.\frac{d}{dt}\right|_{t=0}(\Pi_1
[\gamma_t;1-n]\Pi_1[\gamma_t;1-n]^*)\right)\nonumber
\\
=&\Tr\Biggl((\Pi_1[\gamma;1-n]^*)^{-1}\Pi_1[\gamma;1-n]^{-1}\left(\left.\frac{d}{dt}\right|_{t=0}\Pi_1
[\gamma_t;1-n]\right)\Pi_1[\gamma;1-n]^*\nonumber\\&\hspace{1cm}+(\Pi_1[\gamma;1-n]^*)^{-1}
\left(\left.\frac{d}{dt}\right|_{t=0}\Pi_1
[\gamma_t;1-n]^*\right)\Biggr)\nonumber.
\end{align}Let $\mathrm{u}_t=\gamma_t\circ\gamma^{-1}$ and
\begin{align*}
\mathrm{\dot{u}}(z)=\left.\frac{d\mathrm{u}_t}{dt}\right|_{t=0}(z)=\sum_{k\in
\Z} c_k z^{k+1}.
\end{align*}By \eqref{eq69}, we have
\begin{align*}
\Pi[\gamma_t; 1-n]=\Pi[\gamma;1-n]\Pi[\mathrm{u}_t; 1-n].
\end{align*}Therefore,
\begin{align}\label{eq70}
\left.\frac{d}{dt}\right|_{t=0}\Pi[\gamma_t;
1-n]=\Pi[\gamma;1-n]\left.\frac{d}{dt}\right|_{t=0}\Pi[\mathrm{u}_t;
1-n].
\end{align}By definition,
\begin{align*}
c[1-n]_k \mathrm{u}_t(z)^{k+n-1}
\mathrm{u}_t'(z)^{1-n}=\sum_{l\in\Z}
\Pi[\mathrm{u}_t;1-n]_{lk}c[1-n]_l z^{l+n-1}, \hspace{1cm}z\in S^1.
\end{align*}Differentiate both sides with respect to $t$, we have
\begin{align*}
&\sum_{l\in\Z}\left(
\left.\frac{d}{dt}\right|_{t=0}\Pi[\mathrm{u}_t;1-n]_{lk}\right)c[1-n]_l
z^{l+n-1}\\=&c[1-n]_k \left((k+n-1)
z^{k+n-2}\dot{\mathrm{u}}(z)+(1-n)z^{k+n-1}\dot{\mathrm{u}}'(z)\right)\\
=&c[1-n]_k\sum_{l\in \Z} (k+n-1+ (1-n)(l+1))c_l   z^{l+k+n-1}.
\end{align*}Comparing both sides, we find that
\begin{align*}
\left.\frac{d}{dt}\right|_{t=0}\Pi[\mathrm{u}_t;1-n]_{lk}=&
\frac{c[1-n]_k}{c[1-n]_l}(
k+n-1+(1-n)(l-k+1))c_{l-k}\\=&\frac{1}{\alpha_n}c[1-n]_kc[n]_l(
nk+(1-n)l)c_{l-k}.
\end{align*}From \eqref{eq70} again, we have
\begin{align*}
\left.\frac{d}{dt}\right|_{t=0}\Pi_1 [\gamma_t;1-n]=&\Pi_1[\gamma;
1-n]\left.\frac{d}{dt}\right|_{t=0}\Pi_1
[\mathrm{u}_t;1-n]+\hat{\Pi}_3[\gamma;
1-n]\left.\frac{d}{dt}\right|_{t=0}\hat{\Pi}_2 [\mathrm{u}_t;1-n].
\end{align*}Therefore, \eqref{eq701} is equal to
\begin{align*}
&\Tr\Biggl((\Pi_1
[\gamma;1-n]^*)^{-1}\left(\left.\frac{d}{dt}\right|_{t=0}\Pi_1
[\mathrm{u}_t;1-n]+\left.\frac{d}{dt}\right|_{t=0}\Pi_1
[\mathrm{u}_t;1-n]^*\right)\Pi_1
[\gamma;1-n]^*\\
&+(\Pi_1
[\gamma;1-n]^*)^{-1}\Pi_1[\gamma;1-n]^{-1}\hat{\Pi}_3[\gamma;
1-n]\left(\left.\frac{d}{dt}\right|_{t=0}\hat{\Pi}_2
[\mathrm{u}_t;1-n]\right)\Pi_1
[\gamma;1-n]^*\\
&+(\Pi_1
[\gamma;1-n]^*)^{-1}\left(\left.\frac{d}{dt}\right|_{t=0}\hat{\Pi}_2
[\mathrm{u}_t;1-n]^*\right)\hat{\Pi}_3[\gamma; 1-n]^*\Biggr)\\
=&\Tr\Biggl(\left.\frac{d}{dt}\right|_{t=0}\Pi_1
[\mathrm{u}_t;1-n]+\left.\frac{d}{dt}\right|_{t=0}\Pi_1
[\mathrm{u}_t;1-n]^*\\
&+(\Pi_1[\gamma;1-n]^{-1}\hat{\Pi}_3[\gamma;
1-n]\left(\left.\frac{d}{dt}\right|_{t=0}\hat{\Pi}_2 [\mathrm{u}_t;1-n]\right)\\
&+\left(\left.\frac{d}{dt}\right|_{t=0}\hat{\Pi}_2
[\mathrm{u}_t;1-n]^*\right)\hat{\Pi}_3[\gamma; 1-n]^*(\Pi_1
[\gamma;1-n]^*)^{-1}\Biggr).
\end{align*}Now, since $c_0$ is purely imaginary, and for any $k$,
\begin{align*}
\left.\frac{d}{dt}\right|_{t=0}\Pi_1 [\mathrm{u}_t;1-n]_{kk}=kc_0,
\end{align*}we have
\begin{align*}
\Tr\left(\left.\frac{d}{dt}\right|_{t=0}\Pi_1
[\gamma_t;1-n]+\left.\frac{d}{dt}\right|_{t=0}\Pi_1
[\gamma_t;1-n]^*\right)=0.
\end{align*}On the other hand, by Lemma \ref{lemma11} and Proposition \ref{pro4},
\begin{align*}
\hat{ \Pi}_3[\gamma; 1-n]^*\left(\Pi_1[\gamma;1-n]^*\right)^{-1}
=&\hat{\Pi}_2[\gamma^{-1};
n]\Pi_1[\gamma^{-1};n]^{-1}=\hat{B}[\gamma;n].
\end{align*}Therefore,
\begin{align*}
&\Tr\left(\left(\left.\frac{d}{dt}\right|_{t=0}\hat{\Pi}_2
[\mathrm{u}_t;1-n]^*\right)\hat{\Pi}_3[\gamma; 1-n]^*(\Pi_1
[\gamma;1-n]^*)^{-1}\right)\\
=& \sum_{l=1-n}^{\infty}\sum_{k=n}^{\infty}
B[\gamma;n]_{lk}\left(\left.\frac{d}{dt}\right|_{t=0}\ov{\hat{\Pi}_2
[\mathrm{u}_t;1-n]}_{lk}\right)\\
=&\frac{1}{\alpha_n} \sum_{l=1-n}^{\infty}\sum_{k=n}^{\infty}
 B[\gamma;n]_{lk}c[1-n]_kc[n]_l (nk+(n-1)l)\ov{c_{-l-k}}\\
=&-\frac{1}{\alpha_n} \sum_{l=1-n}^{\infty}\sum_{k=n}^{\infty}
 B[\gamma;n]_{lk}c[1-n]_kc[n]_l(nk+(n-1)l) c_{l+k}\\
=&-\frac{1}{2\pi i}\oint_{S^1} T(z)\tilde{ \mathrm{v}}(z) dz,
\end{align*}where
 $$\tilde{\mathrm{v}}(z) = \sum_{k=1}^{\infty} c_k
z^{k+1},$$and
\begin{align*}
T(z) = &\frac{1}{\alpha_n}\sum_{l=1-n}^{\infty}\sum_{k=n}^{\infty}
B[\gamma;n]_{lk}c[1-n]_kc[n]_l(nk+(n-1)l)z^{-l-k-2}\\
=&-\lim_{w\rightarrow z} \left(\left( n\frac{\pa}{\pa z}
+(n-1)\frac{\pa}{\pa w}\right)\left[\frac{1}{
\alpha_n}\sum_{l=1-n}^{\infty}\sum_{k=n}^{\infty}
B[\gamma;n]_{lk}c[1-n]_kc[n]_lw^{-l-n}z^{-k+n-1}\right]\right)\\
=&-\lim_{w\rightarrow z} \left(\left( n\frac{\pa}{\pa z}
+(n-1)\frac{\pa}{\pa
w}\right)\left[\frac{g'(z)^{1-n}g'(w)^n}{g(z)-g(w)}-\frac{1}{z-w}\right]\right)\\
=&\frac{6n^2-6n+1}{6}\mathcal{S}(g)(z).
\end{align*}Since $\mathcal{S}(g)(z)= O(z^{-4})$ as $z\rightarrow \infty$, the $c_1$ term in $\tilde{\mathrm{v}}(z)$
does not contribute to the integral
$$\oint_{S^1}T(z)\tilde{\mathrm{v}}(z)dz.$$ Therefore, we can replace $\tilde{\mathrm{v}}$ by $\mathrm{v}
=\sum_{k=2}^{\infty} c_k z^{k+1}$ and we have shown that
\begin{align*}
&\Tr\left(\left(\left.\frac{d}{dt}\right|_{t=0}\hat{\Pi}_2
[\mathrm{u}_t;1-n]^*\right)\hat{\Pi}_3[\gamma; 1-n]^*(\Pi_1
[\gamma;1-n]^*)^{-1}\right)\\=&-\frac{6n^2-6n+1}{12\pi i}
\oint_{S^1}\mathcal{S}(g)(z)\mathrm{v}(z) dz.
\end{align*} Finally,
\begin{align*}
&\Tr\left(\Pi_1[\gamma;1-n]^{-1}\hat{\Pi}_3[\gamma;
1-n]\left(\left.\frac{d}{dt}\right|_{t=0}\hat{\Pi}_2 [\mathrm{u}_t;1-n]\right)\right)\\
=&\Tr\left(\left(\left(\left.\frac{d}{dt}\right|_{t=0}\hat{\Pi}_2
[\mathrm{u}_t;1-n]^*\right)\hat{\Pi}_3[\gamma; 1-n]^*(\Pi_1
[\gamma;1-n]^*)^{-1}\right)^*\right)\\
=&\ov{\Tr\left(\left(\left.\frac{d}{dt}\right|_{t=0}\hat{\Pi}_2
[\mathrm{u}_t;1-n]^*\right)\hat{\Pi}_3[\gamma; 1-n]^*(\Pi_1
[\gamma;1-n]^*)^{-1}\right)}\\
=&\frac{6n^2-6n+1}{12\pi
i}\oint_{S^1}\ov{\mathcal{S}(g)(z)}\bar{\mathrm{v}}(z)d\z.
\end{align*}Combining together prove the assertion of the theorem.

\end{proof}

In \cite{Schiffer3} (see also \cite{TT1}),  the function $S:
\Mob(S^1)\bk\Diff_+(S^1)\rightarrow \R$ defined by
\begin{align*}
S(\gamma)
=\iint\limits_{\Del}\left|\frac{f^{\prime\prime}(z)}{f'(z)}\right|^2d^2z
+\iint\limits_{\Del^*}\left|\frac{g^{\prime\prime}(z)}{g'(z)}\right|^2d^2z-4\pi\log|g'(\infty)|
\end{align*} was shown to satisfy
\begin{align*}
\pa S (\mathrm{v})=i\oint_{S^1}\mathcal{S}(g)(z)\mathrm{v}(z)dz
\end{align*}and
\begin{align*}
\pa\bar{\pa}S(\mathrm{v}, \bar{\mathrm{v}})=\Vert\mathrm{v}\Vert^2.
\end{align*}The later implies that $S$ is a Weil-Petersson potential on $\Mob(S^1)\bk\Diff_+(S^1)$.
 Since $\Pi_1[\id; n]=\id$ for all $n\in \Z$, $\mathfrak{F}_n(\id)=0$. Together
with the fact that $S(\id)=0$, we conclude that
\begin{corollary}
On $\Mob(S^1)\bk\Diff_+(S^1)$, we have the following:
\begin{itemize}\item[\textbf{I}.] $$\mathfrak{F}_n
=-\frac{6n^2-6n+1}{12\pi}S.$$\item[\textbf{II}.]Universal Index
Theorem  on $\Mob(S^1)\bk\Diff_+(S^1)$ I.
$$\pa\bar{\pa}\mathfrak{F}_n (\mathrm{v},\bar{\mathrm{v}})= -\frac{6n^2-6n+1}{12\pi}\Vert
\mathrm{v}\Vert^2.$$\\\end{itemize}

\end{corollary}This proves our main result.
\begin{theorem}Universal Index
Theorem  on $\Mob(S^1)\bk\Diff_+(S^1)$ II.

\begin{itemize} \item[\textbf{I}.]For every point on
$\Mob(S^1)\bk \Diff_+(S^1)$, the determinant of period matrix of
holomorphic $n$-differentials $N_n$ is related to the Weil-Petersson
potential $S$ by
$$\det
N_n=\exp\left(-\frac{6n^2-6n+1}{12\pi}S\right).$$
 \item[\textbf{II}.]For every point on
$\Mob(S^1)\bk \Diff_+(S^1)$, the determinant of the period matrix of
holomorphic $n$-differentials $N_n$ is related to the period matrix
of holomorphic one forms $N_1$ by$$\det N_n =(\det
N_1)^{6n^2-6n+1}.$$\end{itemize}
\end{theorem}
The item \textbf{II} of the theorem is the universal version of
Mumford's isomorphism \cite{Mum} $\lambda_n =\lambda_1^{\otimes
6n^2-6n+1}$, where $\lambda_n$ is the determinant line bundle of
$n$-tensor power of the vertical cotangent bundle of the fibration
$\mathcal{C}_g\rightarrow\mathcal{M}_g$, i.e. the fibration of the
universal curve over the moduli space of Riemann surfaces of genus
$g$.

On the other hand, we can also interpret our result in terms of the
Bers integral operator.
\begin{theorem}
For every point on $\Mob(S^1)\bk\Diff_+(S^1)$, the Bers integral
operator $K[n]$, $n\geq 1$ satisfies
\begin{align*}
\det \left(K[n]K[n]^*\right)
=\exp\left(-\frac{6n^2-6n+1}{12\pi}S\right)=\left[\det
\left(K[1]K[1]^*\right)\right]^{6n^2-6n+1}.
\end{align*}
\end{theorem}

\vspace{2cm} \noindent \textbf{Acknowledgement}\; This paper grew
out from a long time discussion with L.A. Takhtajan and A. Mcintyre.
I will like to thank them for the ideas they have given me. I will
also like to thank Academy of Sciences Malaysia for funding this
project under the Scientific Advancement Fund Allocation (SAGA) Ref.
No P96c.

\appendix
\section{A Conjecture}

As we mentioned in the introduction, the Bers integral operator
$K[1]$ is a bounded operator with norm less than or equal to one. We
conjecture that for all $n\geq 2$, $K[n]$ is also a bounded operator
of norm less than or equal to one. Below we give a partial support
to this conjecture.

\begin{lemma}
If $\phi\in A_{1,2}(\Del)$, then $\phi^n\in A_{n,2}(\Del)$.
\end{lemma}
\begin{proof}
Since $\phi\in A_{1,2}(\Del)$, by Lemma 1.3 in Chapter 2 of
\cite{TT1}, $$\Vert\phi\Vert_{1,\infty}=\sup_{z\in\Del}
\rho(z)^{-1/2}|\phi(z)| <\infty.$$ Therefore
$$\iint\limits_{\Del} \rho(z)^{1-n}|\phi^n(z)|^2d^2z\leq
\Vert\phi\Vert_{\infty}^{2n-2}\iint\limits_{\Del}|\phi(z)|^2d^2z<\infty.$$
\end{proof}

\begin{lemma}\label{lemma20}
Let $n\geq 2$ be an integer. If $\phi\in A_{1,2}(\Del)$, then
$$\Vert \phi^n\Vert_{n,2}^2\leq \frac{\beta_1\beta_{n-1}}{\beta_n}\Vert
\phi\Vert_{1,2}^2\Vert \phi^{n-1}\Vert_{n-1,2}^2.$$
\end{lemma}

\begin{proof}
Let
\begin{align*}\phi(z)&=\sum_{k=1}^{\infty}\mathrm{a}_k c[1]_kz^{k-1},\\
\phi^{n-1}(z)&=\sum_{k=n-1}^{\infty}\mathrm{b}_kc[n-1]_kz^{k-n+1},\\
\phi^{n}(z)&=\sum_{k=n}^{\infty}\mathrm{c}_k c[n]_kz^{k-n}.
\end{align*}
Then
$$c[n]_k\mathrm{c}_k=\sum_{l=n-1}^{k-1}c[1]_{k-l}c[n-1]_l\mathrm{a}_{k-l}\mathrm{b}_{l}.$$
We have
\begin{align*}
\Vert \phi^n\Vert_{n,2}^2 =&\sum_{k=n}^{\infty} |\mathrm{c}_k|^2
=\sum_{k=n}^{\infty}\frac{1}{c[n]_k^2}
\left|\sum_{l=n-1}^{k-1}c[1]_{k-l}c[n-1]_l\mathrm{a}_{k-l}\mathrm{b}_{l}\right|^2\\
\leq &\sum_{k=n}^{\infty}\frac{1}{c[n]_k^2}
\left(\sum_{l=n-1}^{k-1}c[1]_{k-l}^2c[n-1]_l^2\right)\left(
\sum_{l=n-1}^{k-1}|\mathrm{a}_{k-l}|^2|\mathrm{b}_{l}|^2\right)
\end{align*}
From the identities
$$\frac{\beta_m}{(1-x)^{2m}}=\sum_{k=m}^{\infty}
c[m]_k^2x^{k-m}$$ and
$$\frac{1}{(1-x)^2}\frac{1}{(1-x)^{2n-2}}=\frac{1}{(1-x)^{2n}},$$
we find that
$$\frac{1}{\beta_1\beta_{n-1}}\sum_{l=n-1}^{k-1}c[1]_{k-l}^2c[1-n]_{l}^2=
\frac{1}{\beta}_n c[n]_k^2.$$ Therefore
\begin{align*}
\Vert \phi^n\Vert_{n,2}^2\leq
&\frac{\beta_1\beta_{n-1}}{\beta_n}\sum_{k=n}^{\infty}
\sum_{l=n-1}^{k-1}|\mathrm{a}_{k-l}|^2|\mathrm{b}_{l}|^2\\
=&\frac{\beta_1\beta_{n-1}}{\beta_n}\left(\sum_{k=1}^{\infty}|\mathrm{a}_k|^2\right)
\left(\sum_{l=n-1}^{\infty}|\mathrm{b}_{l}|^2\right)\\
=&\frac{\beta_1\beta_{n-1}}{\beta_n}\Vert \phi\Vert_{1,2}^2\Vert
\phi\Vert_{n-1,2}^2.
\end{align*}
\end{proof}
The following is the main support of our conjecture.
\begin{theorem}Let $n$ be an integer. Then
\begin{align*}
(K[n]K[n]^*)(z,z) =\iint\limits_{\Del^*}
|K[n](z,w)|^2\rho(w)^{1-n}d^2w\leq \Id[n](z,z).
\end{align*}
\end{theorem}

\begin{proof}

Let $\phi_z\in A_{1,2}(\Del^*)$ be defined as
\begin{align*}
\phi_z(w)= K[1](z,w)=\beta_1\frac{f'(z)g'(w)}{(f(z)-g(w))^2}.
\end{align*}
Then $$K[n](z,w) =\beta_n
\left(\frac{f'(z)g'(w)}{(f(z)-g(w))^2}\right)^n=\frac{\beta_n}{\beta_1^n}\phi_z^n(w).$$
We prove the theorem by induction. The case $n=1$ is known to be
true (see \cite{TT1}). For $n\geq 2$, suppose that
$$(K[n-1]K[n-1]^*)(z,z)\leq \Id[n-1](z,z)=
\frac{\beta_{n-1}}{(1-|z|^2)^{2n-2}},$$ then
$$\Vert\phi_z^{n-1}\Vert_{n-1,2}^2=\frac{\beta_1^{2n-2}}{\beta_{n-1}^2}(K[n-1]K[n-1]^*)(z,z)\leq
\frac{\beta_1^{2n-2}}{\beta_{n-1}}\frac{1}{(1-|z|^2)^{2n-2}}.$$
Therefore by Lemma \ref{lemma20},
\begin{align*}
(K[n]K[n]^*)(z,z)=&\frac{\beta_n^2}{\beta_1^{2n}}\Vert
\phi_z^n\Vert_{n,2}^2 \leq
\frac{\beta_n^2}{\beta_1^{2n}}\frac{\beta_1\beta_{n-1}}{\beta_n}\Vert
\phi_z\Vert_{1,2}^2 \Vert \phi_z^{n-1}\Vert_{n-1,2}^2\\\leq &
\frac{\beta_n\beta_{n-1}}{\beta_1^{2n-1}}\frac{\beta_1^{2n-2}\beta_1}{\beta_{n-1}}\frac{1}{(1-|z|^2)^{2n}}
\\=&\frac{\beta_n}{(1-|z|^2)^{2n}}=\Id[n](z,z).
\end{align*}
By induction, the theorem is true for all $n\geq 1$.
\end{proof}

This theorem implies that the kernel of the operator
$(\Id[n]-K[n]K[n]^*)$ satisfies $(\Id[n]-K[n]K[n]^*)(z,z)\geq 0$.
Therefore, we conjecture that $(\Id[n]-K[n]K[n]^*)$ is a positive
definite operator, which means that $K[n]$ is an operator with norm
less than or equal to one.

\end{document}